\documentclass[%
preprint,
 amsmath,
 amssymb,
 aps,
 prl,
]{revtex4-2}

\usepackage[colorlinks=true,linkcolor=red,citecolor=blue]{hyperref}
\usepackage{graphicx}% Include figure files
\usepackage{dcolumn}% Align table columns on decimal point
\usepackage{bm}% bold math
\usepackage[normalem]{ulem}
\newcommand{\stkout}[1]{\ifmmode\text{\sout{\ensuremath{#1}}}\else\sout{#1}\fi}

\begin{document}
\preprint{Accepted for publication in Soft Matter (RSC)}
\title{Anisotropic short-range attractions precisely model branched erythrocyte aggregates}% Force line breaks with \\
%\thanks{Footnote to title.}%
%
\author{Megha Yadav}
\author{Vanshika}
\author{Chamkor Singh}
\email{chamkor.singh@cup.edu.in}
\affiliation{Department of Physics, Central University of Punjab, Bathinda 151401, India}

\date{\today}% It is always \today, today,
             %  but any date may be explicitly specified
%
\begin{abstract}
Homogeneous suspensions of red blood cells (RBCs or erythrocytes) in blood plasma are unstable in the absence of driving forces and form elongated stacks, called rouleau. These erythrocyte aggregates are often branched porous networks -- a feature that existing red blood cell aggregation models and simulations fail to predict exactly.
Here we establish that alignment-dependent attractive forces in a system of dimers can precisely generate branched structures similar to RBC aggregates observed under a microscope.
Our simulations consistently predict that the growth rate of typical mean rouleau size remains sub-linear -- a hallmark from past studies -- which we also confirm by deriving a reaction kernel taking into account appropriate collision cross-section, approach velocities, and an area-dependent sticking probability.
The system exhibits unique features such as the existence of percolated and/or single giant cluster states, multiple coexisting mass-size scalings, and transition to a branched phase upon fine-tuning of model parameters. Upon decreasing the depletion thickness we find that the percolation threshold increases and the morphology of the structures opens up towards an increased degree of branching. Remarkably the system self-organizes to produce a universal power-law size distribution scaling irrespective of the model parameters.
\end{abstract}
%
%\keywords{Suggested keywords}%Use showkeys class option if keyword
%
\maketitle
%
%%%MAIN TEXT%%%%
\section{Introduction}
Red blood cells (RBCs or erythrocytes) make up about $40-45\%$ by volume of the healthy human blood. In the absence of driving stresses, RBCs aggregate to form stacks, called rouleau -- a feature which plays an important role in rheological properties of blood, in processes like thrombosis and in pathologies such as sickle cell disease, diabetes mellitus, cardiovascular disease, sepsis, and atherosclerosis~\cite{litvinov2017role,satoh1984increased,sheremet2019red}. Furthermore, erythrocyte aggregation is amplified in diseases such as pelvic inflammatory disease~\cite{almog2005enhanced}, obstructive sleep apnea syndrome~\cite{peled2008increased}, and burn injury progression~\cite{clark2018blood}. 
Factors that influence red blood cell aggregation (RBCA) are RBC membrane dynamics, inner cytoskeleton, concentration of plasma protein such as fibrinogen, plasma viscosity, and temperature~\cite{baskurt2011red}. Flow conditions, cell deformability, and shape asymmetry also play a role. Understanding RBCA can significantly help in designing therapeutic targets for the above mentioned diseases.

\begin{figure*}[t!]
	\centering
	\includegraphics[width=0.42\linewidth,angle=-90]{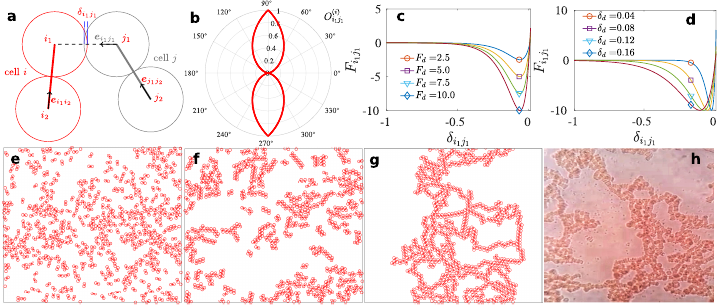}	
	\caption{(a) Schematic of particle 1 of dimer cell $i$ in contact with particle 1 of dimer cell $j$. 
		(b) The alignment factor $O_{i_{1}j_{1}}^{(i)}$ from Eq.~\ref{eq_orientation_factor} as a function of the angle between $\boldsymbol{e}_{i_{1}i_{2}}$ and $\boldsymbol{e}_{i_{1}j_{1}}$. Similarly one can imagine the alignment factor $O_{i_{1}j_{1}}^{(j)}$ from Eq.~\ref{eq_orientation_factor} as a function of the angle between $\boldsymbol{e}_{j_{1}j_{2}}$ and $\boldsymbol{e}_{i_{1}j_{1}}$.
		(c-d) The depletion force $F_{i_{1}j_{1}}^\mathrm{d}$ [Eq.~\ref{eq_fd}] between particle $1$ of dimer $i$ and  particle $1$ of dimer $j$ as a function of the overlap $\delta_{i_{1}j_{1}}$ with varying strength of depletion force $F_d$, and varying depletion thickness or range $\delta_d$, respectively. Negative $\delta_{i_{1}j_{1}}$ means the particles are separated. 
		(e-g) The transition from a distributed non-branched phase towards a branched and percolated phase. The model parameters are $t=1200$, $n=0.25$, $\delta_d=0.04$, and the ratio of the depletion force strength to the thermal force strength $F_d/F_T$ increases from (e) to (g). 
		(h) A lab microscope (ECLIPSE E200) image of human erythrocytes forming a branched porous network, very similar to the one simulated in (g).
	}
	\label{fig_model}
\end{figure*}

RBCA is driven by surface macromolecule bridging~\cite{chien1973ultrastructural} and depletion of macromolecule concentration in the surrounding liquid medium~\cite{wagner2013aggregation}; these two remain competing theories of RBCA~\cite{baskurt2011red}. The depletion interaction between RBCs is experimentally well characterized~\cite{steffen2013quantification}. %
Shapes of RBCs within rouleaux as a function of macromolecular concentration using confocal microscopy have been studied~\cite{flormann2017buckling} and it is known that different configurations might exist in equilibrium, and deformation increases non-linearly with interaction energy or force. 
Different physical aspects {\it e.g.} effect of surface macromolecules, rheology as a function of macromolecule adsorption rate, yield stress of RBC aggregates, sedimentation rates, and aggregation index have been explored~\cite{flormann2017physical}. Strong dependence of morphology of interaction zones on macromolecule concentration is there and the viscosity of the surrounding medium adds to the complexity of RBCA~\cite{flormann2017physical}. The morphology alters rheological behavior such as shear thinning~\cite{lanotte2016red}. 
Experiments have been used to test theoretical models and assumptions by visualizing RBCA in plasma~\cite{barshtein2000kinetics} and confirm that RBCA involves both polymerization (erythrocyte-erythrocyte aggregation) and condensation (rouleaux-rouleaux aggregation) with nearly same values of rate constants for erythrocyte-erythrocyte, erythrocyte-rouleaux, and rouleaux-rouleaux aggregation~\cite{barshtein2000kinetics}.
On top of this, biological cells may execute Brownian or other type of stochastic motion which sometimes is resisted by surfaces in contact~\cite{sewchand1982resistance,singh2021guided}.
Detailed treatment of determinants of RBCA such as shape, hematocrit, plasma proteins, dextran or polymer concentrations; mechanisms of RBCA such as bridging of macromolecules on the surface of RBCs or depletion of macromolecules in the plasma; and RBCA measurement techniques are compiled in seminal work of~\citet{baskurt2011red}.

One of the earliest attempts to physically describe erythrocyte aggregation and sedimentation dynamics was made by Ponder~\cite{ponder1925sedimentation,ponder1926sedimentation,ponder1932sedimentation} in a series of experimental and theoretical studies.
\citet{ponder1926sedimentation} predicted linear growth using Smoluchowski's approach, and compared the result with experiments for short times. The Smoluchowski approach used by \citet{ponder1926sedimentation} was unable to predict the sublinear growth later in time, perhaps partly due to the fact that appropriate sticking probability, approach velocities, and collision cross sections were not taken into account. 
\citet{kernick1973experiments} experimentally showed that the mean rouleaux size increases linearly with time during the early phase of aggregation, and after the initial phase, the mean size versus time curve flattens.
Simple kinetic models to understand linear and branched rouleau formation were proposed by ~\citet{samsel1982kinetics} improving upon the work of \citet{ponder1926sedimentation} to predict the size distribution of rouleau. The model was then extended to include dissociation/fragmentation as well~\cite{samsel1984kinetics}. Models are also proposed for equilibrium size distributions of RBC aggregates~\cite{perelson1982equilibrium}.
Ponder's~\cite{ponder1926sedimentation} application of the Smoluchowski equation of colloid aggregation assumes that the process is irreversible. But rouleaux are not permanent aggregates. Both aggregation and dissociation may go on in a statistically steady state manner and the model should incorporate such dynamics. The result of linear growth of mean rouleau size is thus not universal. 

Average size and growth rates in RBCA have been studied extensively, for example, using optical transmittance~\cite{bertoluzzo1999kinetic} to explore the effects of temperature, plasma dilutions, different macromolecular solutions, and membrane alterations using alkylating agents.
{In addition, configurations and deformations of two joining erythrocyte cells have proven to be sensitive to the interaction parameters, {\it e.g.} the range of attraction~\cite{babaki2023competition}.} 
Physically, in the recent, it has come to the knowledge that purely hydrodynamically interacting particles in pressure-driven flow exhibit universal scaling of the cluster size distribution independent of the concentration and particle shape~\cite{ding2006cluster}.
Also, RBCs under gravity sediment and may collapse to form a soft particle gel at high volume concentrations~\cite{darras2022erythrocyte}. The motion of RBCs is dependent upon intricate interaction between confinement and deformation~\cite{olla1999simplified}. On top of flow-dependent shape instability, shape asymmetry, and shape transitions~\cite{kaoui2009red,mauer2018flow}, the membrane dynamics of red blood cells is an intricate subject~\cite{gov2003cytoskeleton,rochal2006cytoskeleton,sens2007force,boal1992dual,ben2011effective,pivkin2008accurate}.  
RBCs under shear exhibit sub-diffusive motion~\cite{grandchamp2013lift} and may assemble into crystal-like patterns in a confined shear flow~\cite{shen2018blood}. Different motion modes like swinging~\cite{abkarian2007swinging}, chaotic motion under cyclic shear~\cite{dupire2010chaotic}, tank-treading and tumbling~\cite{skotheim2007red} have been observed, and distinct configurations appear in erythrocyte-erythrocyte doublet dynamics~\cite{abbasi2021erythrocyte}. If the rates of RBCA can be predicted precisely, then we can understand processes like immune-haemagglutination -- the attachment of antibodies in the blood plasma to antigen molecules at the surface of (foreign) RBCs -- in which the number of free erythrocytes initially decrease exponentially~\cite{ming1965mathematical} but long time dynamics is still unclear.  

Despite numerous studies on above mentioned complex aspects of RBCA, one aspect that remains quantitatively unexplored is that RBCs undergo a transition to gel-like branched and porous matrix structures, observable under a microscope. \citet{fenech2009particle} proposed a spherical particle model to study RBCA in an attempt to link macroscopic blood properties with micromechanical cell interactions. 
More recently \citet{nehring2018morphology} constructed a more involved RBC model by bonding multiple spherical beads together ($19$ beads in one RBC) and were able to simulate face-to-face and face-to-side aggregate structures in small system sizes of $\sim10$ RBCs at low volume fraction limit. Although mesoscale aspects of the RBC aggregate morphology were well captured, long rouleaux formation and complex network formation with multiple branching nodes still remain unexplored. Moreover, larger system sizes require an efficient model scheme such that the simulation times can be reduced, while at the same time, not losing essential aspects of the RBC aggregate network morphology. Currently, available models and simulations do not capture this feature of RBCA in an exact manner. In this study, we make a proposition that a system of dimers with alignment-dependent attractive forces can precisely generate branched porous structures and mimic erythrocyte networks observed under a lab microscope. Here by "precise" we mean that the model consistently predicts the sub-linear growth observed in experiments, is able to reproduce branching upon fine tuning of model parameters, and also predicts a universal power law size distribution independent of model parameters, which sometimes is expected in self-organizing systems~\cite{bak1988self,bak59self}. Most remarkably, we find that the degree of branching increases upon decreasing the depletion thickness. We find percolated and single-cluster states on a master curve between the average size and a non-dimensional parameter involving depletion force strength, depletion thickness, temperature, and number density. Finally, we show -- by constructing an appropriate collision/reaction kernel -- that Smoluchowski's aggregation equation endorses the sublinear growth regime predicted by the simulation model, if the effects of the geometric shape of the rouleaux, their approach velocities, and appropriate sticking probabilities are taken into account in the collision/reaction kernel.

\section{Model}
A dimer composed of two round particles each having unit size ($d=1$), and held together by a linear spring and a dashpot, is used to represent the RBC shape in quasi two dimensions [Fig.~\ref{fig_model} (a)]. An alignment-dependent attractive force, representative of the protein depletion interaction, between particles belonging to two different dimers -- say the force exerted by particle 1 of dimer cell $j$ on particle 1 of dimer cell $i$ -- is implemented as [Fig.~\ref{fig_model} (a-d)]    
\begin{align}
	\boldsymbol{F}_{i_{1}j_{1}}^\mathrm{d}=
	\Theta(r_c-|\boldsymbol{r}_{i_1}-\boldsymbol{r}_{j_1}|) \; F_{i_{1}j_{1}}^\mathrm{d} \; O_{i_{1}j_{1}} \:
	\boldsymbol{e}_{i_{1}j_{1}} ,
	\label{eq_fd_1}
\end{align}
where $\Theta$ is the Heaviside step function with cutoff radius $r_c=2d$, $ \boldsymbol{e}_{i_{1}j_{1}}=(\boldsymbol{r}_{i_1}-\boldsymbol{r}_{j_1})/
{|\boldsymbol{r}_{i_1}-\boldsymbol{r}_{j_1}|}$ is a unit vector pointing from the center of particle 1 of dimer cell $j$ towards the center of particle 1 of dimer cell $i$, $F_{i_{1}j_{1}}^\mathrm{d}$ is given by    
\begin{align}
	F_{i_{1}j_{1}}^\mathrm{d}=-4F_d\left(
	\exp \left[\frac{(\delta_{i_{1}j_{1}}-\delta_o)}{\delta_d} \right]
	-\exp \left[\frac{2(\delta_{i_{1}j_{1}}-\delta_o)}{\delta_d} \right]
	\right),
	\label{eq_fd}
\end{align}
and the alignment factor
\begin{align}\nonumber
	O_{i_{1}j_{1}}&=(1-|\boldsymbol{e}_{i_{1}j_{1}}\cdot\boldsymbol{e}_{i_{1}i_{2}}|)\times(1-|\boldsymbol{e}_{i_{1}j_{1}}\cdot\boldsymbol{e}_{j_{1}j_{2}}|)\\
	&\equiv O_{i_{1}j_{1}}^{(i)}\times O_{i_{1}j_{1}}^{(j)}.
	\label{eq_orientation_factor}
\end{align}
Here $F_d$ is the strength of the depletion force, $\delta_{i_{1}j_{1}}\equiv d- |\boldsymbol{r}_{i_1}-\boldsymbol{r}_{j_1}|$ is the overlap between particle 1 of cell $i$ and particle 1 of cell $j$, $\delta_o$ is equilibrium overlap, and $\delta_d$ is the range of the force representing depletion thickness in the actual RBC system.
\begin{figure*}[t!]
	\centering
	\includegraphics[width=1.0\linewidth]{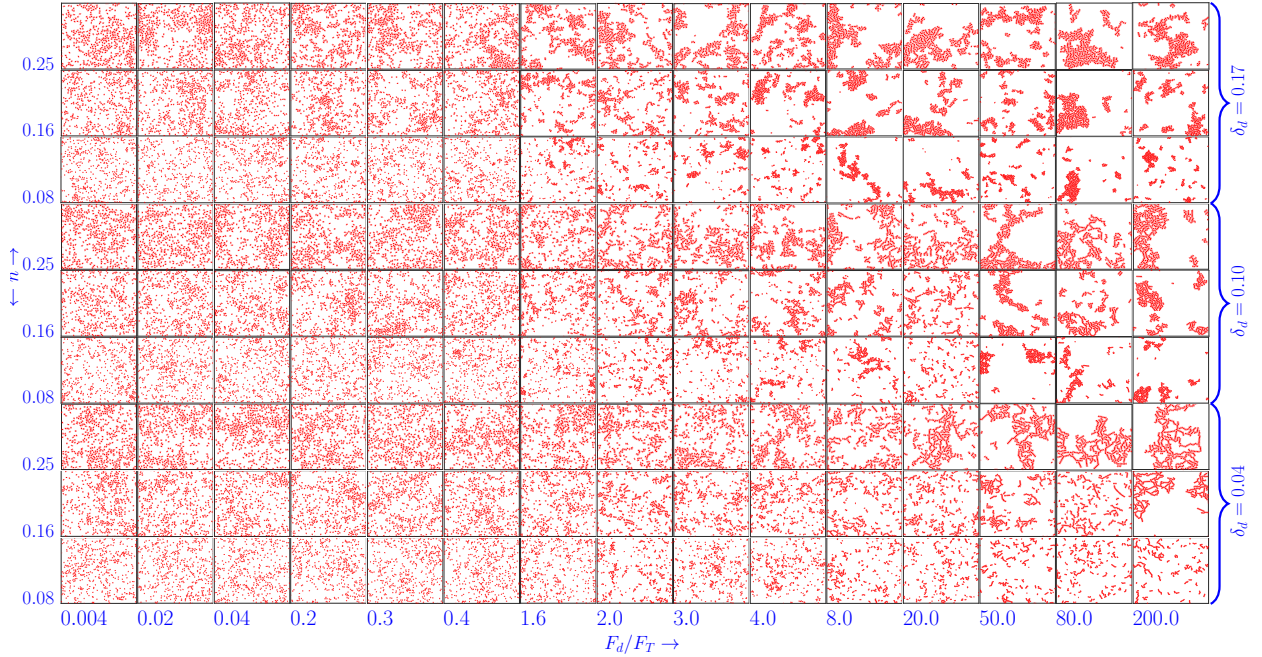}	
	\caption{ {A compilation of configurations from a set of 135 simulations at different values of the number density $n$, depletion thickness $\delta_d$ signifying the range of the attractive depletion force [Eq.~\ref{eq_fd}], and the ratio of the depletion force strength to the thermal force strength $F_d/F_T$.} Each configuration is taken at time $t=1200$ when a statistically stationary (but still dynamic) state has been reached. Clearly at high $F_d/F_T$ -- for instance in the last three columns -- the system transitions towards a branched porous configuration from a more dense configuration upon decreasing the depletion range $\delta_d$ at relatively high number density.}
	\label{fig_config}
\end{figure*}
The depletion force as a function of the separation, without the effect of the alignment factor, is shown in Fig.~\ref{fig_model} (c, d) for varying $F_d$ and $\delta_d$.
\begin{figure*}[t!]
	\centering
	\includegraphics[width=0.85\linewidth]{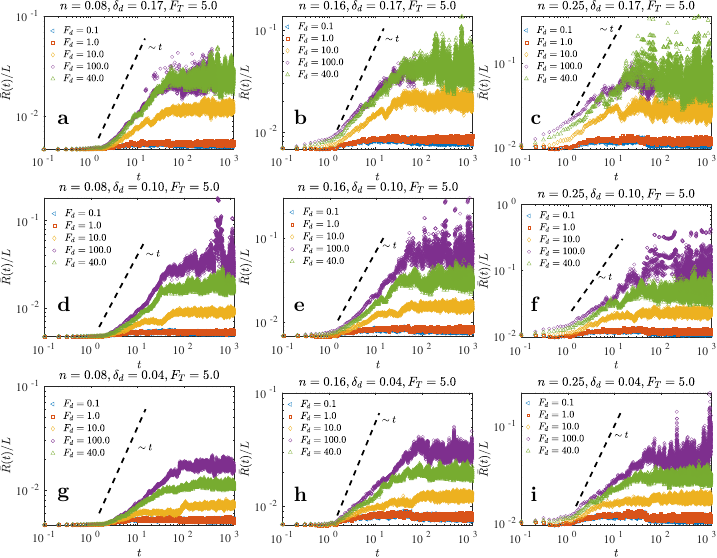}	
	\caption{Mean radius of gyration $\bar{R}(t)$ of aggregates, scaled by the system size $L$, versus time under varying number density $n$, depletion thickness $\delta_d$ and depletion force strength $F_d$, at $F_T=5.0$. The growth consistently remains sublinear. After the growth period, a statistically stationary regime is reached where $\bar{R}(t)$ fluctuates about a steady mean signifying the dynamic aggregation and dissociation/fragmentation events. Each curve is one simulation.}
	\label{fig_R_vs_t}
\end{figure*}
Although there are a number of interaction potentials that can be used to model attractions, we choose the specific force model in Eq.~\ref{eq_fd} because we can fine-tune the strength as well as the range of the depletion interaction using only two parameters: $F_d$ and $\delta_d$. The model is testable using techniques, such as optical tweezers~\cite{lee2016optical}. The dimer model has another advantage that we can neglect the rotational degrees of freedom of individual sub-particles and still the dimers can have rotational degrees of freedom through different translations executed by its sub-particles.
Additional aspect of our simulations is the alignment factor $O_{i_{1}j_{1}}\equiv O_{i_{1}j_{1}}^{(i)}\times O_{i_{1}j_{1}}^{(j)}$ as a function of the angles between unit vectors $\boldsymbol{e}_{i_{1}j_{1}}$ and $\boldsymbol{e}_{i_{1}i_{2}}$, and $\boldsymbol{e}_{i_{1}j_{1}}$ and $\boldsymbol{e}_{j_{1}j_{2}}$. The first part $O_{i_{1}j_{1}}^{(i)}=(1-|\boldsymbol{e}_{i_{1}j_{1}}\cdot\boldsymbol{e}_{i_{1}i_{2}}|)$ is shown in Fig.~\ref{fig_model} (b). 
The factor $O_{i_{1}j_{1}}$ mimics the alignment-dependent depletion-based attractions between RBCs. %
One reason behind designing the alignment factor $O_{i_{1}j_{1}}$ for anisotropic attractions between RBCs is that the alignment-dependent attractions are able to cleanly generate {\it sharp} branch nodes. On the other hand, only distance dependence attractions generate relatively {\it thicker} nodes.
To avoid excessive overlap between cells and to mimic physical dissipation, a standard Hertzian spring and linear dashpot starts acting if $\delta_{i_{1}j_{1}}>0$, written as   
\begin{equation}
	\boldsymbol{F}_{i_1j_1}^\mathrm{contact} =  \Theta(\delta_{i_1j_1})\:
	\: 
	\left[k \delta_{i_1j_1}^{3/2}
	- \gamma
	\: \boldsymbol{v}_{i_1j_1}\cdot\boldsymbol{e}_{i_1j_1} \right] 
	\:
	\boldsymbol{e}_{i_1j_1},
	\label{eq_contact}
\end{equation}
where $\boldsymbol{e}_{i_1j_1} =(\boldsymbol{r}_{i_1}-\boldsymbol{r}_{j_1})/
{|\boldsymbol{r}_{i_1}-\boldsymbol{r}_{j_1}|}$, $ \boldsymbol{v}_{i_1j_1} = \boldsymbol{v}_{i_1}-\boldsymbol{v}_{j_1}$, $\Theta$ is the Heaviside step function, and $k,\:\gamma$ are elastic and damping constants. 
{Although the RBC is assumed to have a shape made up of two rigid circles, the ``hardness” or ``softness” of these circles can still effectively be tuned by changing the elastic and damping constants in the model. Setting a low elastic constant results in relatively higher overlap between touching particles which effectively models the deformation and thus the ``softness” of RBCs. In our simulations, the elastic constant is not that high (see Appendix I) and thus it is plausible to assume that the model effectively takes into account at least the zeroth order aspects of deformations of the RBCs.}
In addition, to replicate the temperature of the plasma surrounding the RBCs, a stochastic thermal force acts on each particle. %
{
	Even though the thermal energy in the context of RBCs can be considered negligible, there exist past experiments which have reported certain interesting effects of temperature on the RBC aggregation, such as temperature for optimal growth of aggregates~\cite{kernick1973experiments}.} We add this force on particle $i_1$ as
\begin{equation}
	\boldsymbol{F}_{i_1}^\mathrm{thermal} =  F_T\;\boldsymbol{\eta}_{i_1}(t),
	\label{eq_ft}
\end{equation}
where $F_T$ is the strength of the thermal force and $\boldsymbol{\eta}_{i_1}(t)\equiv({\eta}_{i_1,x}(t),\;{\eta}_{i_1,y}(t))$ is a vector whose components ${\eta}_{i_1,x}(t),\;{\eta}_{i_1,y}(t)$ at time instant $t$ are randomly chosen from a parent normal distribution with zero mean and unit standard deviation. The thermal force is delta-correlated in time, or in other words, it has a flat power spectrum density. 
Finally, the total external force on particle $1$ of cell $i$ exerted by particle $1$ of cell $j$ is
\begin{equation}
	\boldsymbol{F}_{i_1}^\mathrm{total} = \boldsymbol{F}_{i_1j_1}^\mathrm{d} + \boldsymbol{F}_{i_1j_1}^\mathrm{contact} +\boldsymbol{F}_{i_1}^\mathrm{thermal}+
	\boldsymbol{F}_{i_1i_2}^\mathrm{internal},
\end{equation}
and the same process then can be extended to all other interacting particle pairs. Note that the force $\boldsymbol{F}_{i_1i_2}^\mathrm{internal}$  is due to the fact that the sub-particles belonging to a given dimer exert forces on each other as well, as they are held together by a linear/harmonic spring and dashpot. Once the forces are calculated, one can integrate the equations of motion; details of the numerical scheme are described in Appendix I.
\begin{figure*}[t!]
	\centering
	\includegraphics[width=0.833\linewidth]{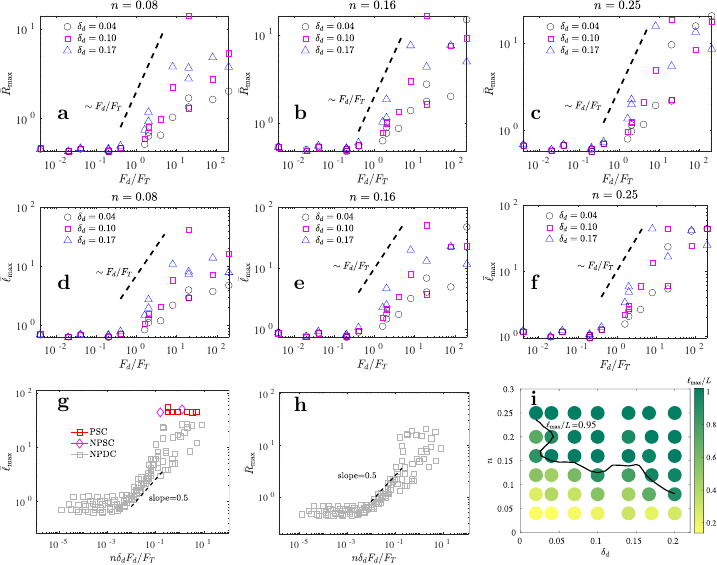}
	\caption{(a-f) Maximum of the mean radius of gyration ($\bar{R}_\mathrm{max}\equiv \max [\bar{R}(t)]$) and the maximum of the mean linear size ($\bar{\ell}_\mathrm{max}\equiv \max [\bar{\ell}(t)]$), versus the ratio of the depletion force strength to the thermal force strength $F_d/F_T$, for varying number density $n$ and varying depletion thickness $\delta_d$.  {Aggregation starts approximately as ${F_d}/{F_T} \rightarrow 1$.} Each data point represents one simulation.
		{(g-h) The rescaled data using single non-dimensional group $n\delta_dF_d/F_T$. Here PSC means that the the system contains a percolating single cluster, NPSC means non-percolating single cluster and NPDC means non-percolating distributed clusters. {After the initiation of aggregation at ${F_d}/{F_T} \approx 1$, a scaling of the form $\bar{\ell}_\mathrm{max}\sim (n\delta_dF_d/F_T)^{1/2}$ governs an intermediate growth regime up to ${n\delta_dF_d}/{F_T} < 10^{-1}$, after which the scaling breaks. Its in this regime of ${n\delta_dF_d}/{F_T} > 10^{-1}$ where the PSC or NPSC states appear.} Each data point represents one simulation.
		(i) Estimate of the percolation threshold ({\it i.e.} the critical number density at or above which $\ell_\mathrm{max}\rightarrow L$) as a function of $\delta_d$. The solid line is an interpolated contour level at which $\ell_\mathrm{max}/L=0.95$ to indicate an approximate separation between percolated and non-percolated states. The percolation threshold decreases with increasing depletion range. The parameter $F_d/F_T=80$ is fixed in this case and each colored circle is one simulation.
		}
	}
	\label{fig_ell_vs_f}
\end{figure*}
\section{Scaling, percolation, and universality}
Dimer shape and alignment-dependent depletion forces in the system lead to various organizational states. We tune the four model parameters, namely the number density $n\equiv N/L^2$ {\it i.e.} total number of particles per unit area in two dimensions, depletion force strength $F_d$ and depletion thickness $\delta_d$ [Eq.~\ref{eq_fd}, Fig.~\ref{fig_model} (c-d)], and the strength of the thermal force $F_T$ [Eq.~\ref{eq_ft}]. { The number density is changed by changing the size of the simulation box. This way we keep particle number $N$ fixed for all simulations.} After some initial time period in which the aggregates nucleate from a nearly homogeneously dispersed state, the process proceeds to a growth period, and then to a dynamic quasi-steady state where the aggregates form and fragment continuously. As a typical example of change in the organization, a distributed aggregate state transitions to a percolated and highly branched state upon increasing the ratio of the depletion force strength to the thermal force strength $F_d/F_T$ at fixed number density $n=0.25$, depletion thickness $\delta_d=0.04$, and at time $t=1200$ [Fig~\ref{fig_model} (e-g)]. These percolated and branched RBC states we also observe under a lab microscope [Fig~\ref{fig_model} (h)]. 
In a distributed state, aggregates continuously form and break due to competition between depletion and thermal forces, while in a percolated state there is at least one cluster that spans the system. A third scenario is the emergence of a single (percolating or non-percolating) giant cluster where all the cells aggregate together. 
Typical configurations during dynamic quasi-steady state at time $t=1200$ for different combinations of the ratio of depletion to thermal force strength $F_d/F_T$, number density $n$, and depletion thickness $\delta_d$ are compiled in Fig.~\ref{fig_config}. 

To study the structures quantitatively, we differentiate aggregates based on the condition that if two particles are in contact, they belong to the same aggregate. {In the following we use the term "aggregate" or "cluster" interchangeably. For simplicity the persistence time of a contact is not taken into the definition of an aggregate.} At any given time $t$, we measure the size of $k^{th}$ aggregate using two statistical quantities. First is the radius of gyration ${R}_k(t) \equiv [\sum_{i,j} |\boldsymbol{r}_i(t)-\boldsymbol{r}_j(t)|^2/(2N_k^2)]^{1/2}$ where the pairs $i,j$ belong to the $k^{th}$ aggregate having $N_k$ particles in it. 
The second quantity is span of the $k^{th}$ aggregate ${\ell}_k(t)=[\ell_{k,x}(t)+\ell_{k,y}(t)]/2$ where $\ell_{k,x}$ and $\ell_{k,y}$ are horizontal and vertical spans of the $k^{th}$ aggregate respectively. Then at any given time, we find the means $\bar{R}(t)=\sum_k R_k(t)/N^{a}$, $\bar{\ell}(t)=\sum_k \ell_k(t)/N^{a}$ with $N^{a}$ being total number of aggregates present in the system at time $t$. 
{The coordinates of particles in a cluster, which spans the system due to periodic boundary conditions, are boundary corrected before the calculation of ${R}_k(t)$ or ${\ell}_k(t)$.}
Time evolutions of $\bar{R}(t)$ scaled by the system size $L$ are depicted in Fig.~\ref{fig_R_vs_t} for varying parameters. Under favorable conditions for any considerable growth of aggregates to happen, we observe nearly a power law
\begin{align}
	\bar{R}(t) \sim t^\alpha, \:\alpha<1,	
\end{align}
during the growth period depicting a sublinear scaling of the mean radius of gyration with time. The power law growth is apparent for $F_d/F_T\gg 1$ in Fig.~\ref{fig_R_vs_t}.
After the growth period in which $\bar{R}$ increases with $t$, the system enters a dynamic stage where aggregates form and fragment stochastically with $\bar{R}$ fluctuating with time around a mean. We label this stage as a statistically stationary state with fluctuating aggregation and fragmentation events. 
{We note that variation of $\bar{R}$ in certain cases is nearly an order of magnitude, and these fluctuations reduce upon decreasing $n$, $\delta_d$ and $F_d/F_T$ in the model.}
Upon increasing the depletion force strength $F_d$, the growth exponent $\alpha$ as well as the steady state $\bar{R}$ value increases. Of particular interest are the maximum attainable values of $\bar{R}(t)$ and $\bar{\ell}(t)$ -- namely $\bar{R}_\mathrm{max}$ and $\bar{\ell}_\mathrm{max}$ respectively -- which are plotted in Fig.~\ref{fig_ell_vs_f} (a-f) as a function of the ratio $F_d/F_T$ for different $n$ and $\delta_d$. 
{Note that although appreciable aggregation starts for ${F_d}/{F_T} > 1$, there still exist some aggregate size fluctuations for weak attractions relative to the thermal force $F_d/F_T < 1$ because the aggregate sizes are calculated instantaneously and the persistence time of contacts is not taken into the definition of an aggregate.}
Considering that the particles in dimer cells are of non-dimensional unit mass and size, and $\bar{\ell}_\mathrm{max}$ -- the biggest peak in a $\bar{\ell}(t)$ curve -- depends on $n,\:\delta_d,\:F_d$, and $F_T$, we posit that the biggest aggregate size would scale as
\begin{align}
	\bar{\ell}_\mathrm{max} \sim n^a \delta_d^b \left[\frac{F_d}{F_T}\right]^c,
	\label{eq_ell_max_scaling}
\end{align}
where $a,\:b,\:c$ are exponents to be determined from the simulations. The simulation data is shown in Fig.~\ref{fig_ell_vs_f} (g).
{After the initiation of aggregation at ${F_d}/{F_T} \approx 1$, the data tend to follow a power law scaling of the form 
	\begin{equation}
		\bar{\ell}_\mathrm{max} \sim \left[\frac{n\delta_d F_d}{F_T}\right]^{1/2}
	\end{equation}
	for an intermediate regime up to $n \delta_d {F_d}/{F_T} < 10^{-1}$ [Fig.~\ref{fig_ell_vs_f} (g)] not only for $\bar{\ell}_\mathrm{max}$ but also for the another measure of the biggest cluster size $\bar{R}_\mathrm{max}$ [Fig.~\ref{fig_ell_vs_f} (h)]. A total of $135$ simulations are used to mark the scaling in this intermediate regime. We posit that an appropriate approximation for exponents in Eq.~\ref{eq_ell_max_scaling} for the intermediate regime is $a=b=c=1/2$. For $n \delta_d {F_d}/{F_T} > 10^{-1}$, the data is somewhat scattered to endorse the above scaling, however, it is in this regime where percolating ($\bar{\ell}_\mathrm{max}\rightarrow L$ or $\bar{\ell}_\mathrm{max}/L\rightarrow 1$ where $L$ is the system size) and/or single cluster states emerge.} This suggests that underlying the growth curves in Fig.~\ref{fig_R_vs_t} and \ref{fig_ell_vs_f}, there exist variations in the morphology of the aggregates. We mark these on the curve in Fig.~\ref{fig_ell_vs_f} (g) as ($i$) percolating single cluster, if $\bar{\ell}_\mathrm{max}=L$ and all the cells have aggregated to a single cluster, ($ii$) non-percolating single cluster, if $\bar{\ell}_\mathrm{max}<L$ but all the cells have aggregated to a single cluster, and ($iii$) non-percolating distributed clusters if neither $\bar{\ell}_\mathrm{max}=L$ nor all the cells have aggregated to a single cluster.
An important observation is that the percolation threshold, {\it i.e.} the critical number density for which $\ell_\mathrm{max}\rightarrow L$, depends on the depletion thickness $\delta_d$. This dependence is depicted in Fig.~\ref{fig_ell_vs_f} (i) for the case $F_d/F_T=80$. If we reduce the range of the depletion force, the aggregates open up however we require a relatively high number density for percolation to happen, as apparent in Fig.~\ref{fig_config}. To mark the percolation threshold more quantitatively, we interpolate a contour level  $\ell_\mathrm{max}/L=0.95$ in Fig.~\ref{fig_ell_vs_f} (i) which approximately separates the percolated and non-percolated states. This level value of less than $1$, but not exactly $1$, separates states that are not percolated from the ones that almost approach percolation. Clearly, the threshold decreases with increasing $\delta_d$.

\begin{figure*}[t!]	
	\centering
	\includegraphics[width=0.575\linewidth, angle=90]{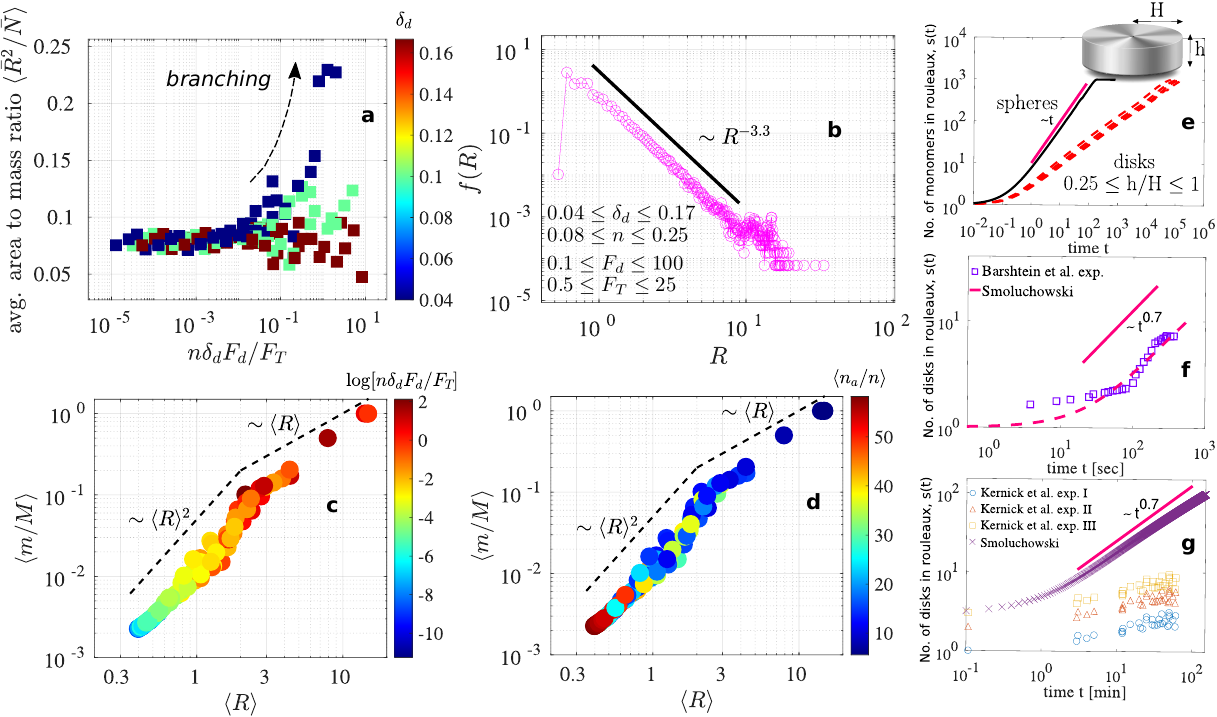}
	\caption{		
		(a) Average area-to-mass ratio of aggregates as a measure of the degree of branching. Branching is amplified upon increasing the parameter $n\delta_dF_d/F_T$ but with decreasing $\delta_d$ (color bar).
		(b) The distribution of sizes of the aggregates which are collected in the time window $1000 \leq t \leq 1200$ during statistically stationary evolution of the system for a total of 135 combinations of $n$, $\delta_d$, $F_d$, and $F_T$.
		{(c-d) Scaled average aggregate mass $\langle m(t)/M\rangle$ versus average radius of gyration $\langle R(t)\rangle$. Here $M$ is the total mass in the system, and $\langle\rangle$ denotes the average over an ensemble collected in the time window $1000 \leq t \leq 1200$ during the statistically stationary evolution of the system. The scaling exponent is the fractal dimension $D_f$: $D_f\approx 2$ for $\langle R\rangle<2$ and there is a signature that $D_f< 2$ for relatively larger aggregates $\langle R\rangle>2$. The color bar in (c) depicts the value of the nondimensional group $n\delta_d F_d/F_T$} corresponding to the data points, and the color bar in (d) depicts how dense or compact the aggregates are for the corresponding data points. See text for the definition of the average relative number density of aggregates $\langle n_a/n\rangle$.
		(e) Solutions~\ref{eq_nt_spheres} and~\ref{eq_nt_disks} of the Smoluchowski's Eq.~\ref{eq_dndt} using the derived reaction kernel Eq.~\ref{eq_K}. (f) Solution for the disks compared with experimental data from~\citet{barshtein2000kinetics}, and (g) compared with experimental data from~\citet{kernick1973experiments}.			
	}
	\label{fig_df}
\end{figure*}

A fundamental observation in our study is that the dense aggregate states, that appear at high $\delta_d$ and high $F_d/F_T$ in Fig.~\ref{fig_config}, transition toward states with more and more porous and branched structures upon decreasing the depletion thickness $\delta_d$. This is clear if we move along decreasing $\delta_d$ in the last three columns in Fig.~\ref{fig_config}. This morphological transition is one of the main observations and we quantitatively characterize it using a couple of methods as follows.
First, the area covered by the structures divided by their masses should increase as the degree of branching within these structures increases and the aggregates turn more and more porous. The average area occupied by the aggregates at a given time is estimated simply by $\bar{R}^2$ and the average mass of aggregates at the same time is proportional to $\bar{N}$ -- the average number of particles in an aggregate. Average over an ensemble of configurations within a time window $1000\leq t \leq 1200$ is then taken, {\it i.e. $\langle\bar{R}^2/\bar{N}\rangle$}. This measure of the degree of branching is shown in Fig.~\ref{fig_df} (a). The plot precisely depicts that on average the area-to-mass ratio of aggregates increases if we increase the value of $n\delta_dF_d/F_T$ but only if we decrease the depletion thickness $\delta_d$ (blue data points in Fig.~\ref{fig_df} (a)). At $\delta_d=0.04$, the average area-to-mass ratio almost diverges as we increase $n\delta_dF_d/F_T$ clearly marking the transition to the branched phase. 
Second, we compute average fractional aggregate mass $\langle m(t)/M\rangle$ and average aggregate size $\langle R(t)\rangle$ where $m(t)$ is the mass of an aggregate, $M$ is the total mass in the system, $R(t)$ is the radius of gyration of the aggregate, and $\langle\rangle$ denotes average over an ensemble of aggregates which appear in a time window $1000 \leq t \leq 1200$ during the statistically stationary evolution. The result is shown in Fig.~\ref{fig_df} (c, d) where $\langle m/M\rangle \sim \langle R\rangle^{D_f}$. The scaling exponent $D_f$ is the fractal dimension: we find $D_f \approx 2$ for $\langle R\rangle<2$ and there is a signature that $D_f<2$ for relatively larger aggregates $\langle R\rangle>2$. A lower value of $D_f$ for larger size aggregates appearing in the system indicates that they are relatively more porous. The color bar in Fig.~\ref{fig_df} (c) depicts the logarithm of the corresponding values of parameter $n\delta_d F_d/F_T$.
For additional measure of how dense or compact the aggregates are, we compute the relative number density of a given $k^{th}$ aggregate $ n_k/n \equiv [N_k/(\pi R_k^2)]/n$ where $N_k$ are the total number of particles in the $k^{th}$ aggregate, $R_k$ is the radius of gyration of $k^{th}$ aggregate, and $n$ is the overall system number density. The ensemble-averaged value of this quantity is depicted on the color bar in Fig.~\ref{fig_df} (d). Thus in a nutshell from Fig.~\ref{fig_df} (c, d) we find that smaller aggregates that form mostly at lower $n\delta_d F_d/F_T$ values are relatively denser or compact with fractal dimension $D_f\approx2$, while larger aggregates forming mostly at higher $n\delta_d F_d/F_T$ values are relatively fluffy or porous with fractal dimension $D_f<2$.
If we collect an ensemble of aggregate radii of gyration $R$ appearing within a time window $1000\leq t \leq 1200$ irrespective of the values of the model parameters $n$, $\delta_d$, $F_d$ and $F_T$, and we plot the size distribution function $f(R)$, we find a universal power law behavior: $f(R)\sim R^{-3.3}$ [Fig.~\ref{fig_df} (b)]. Note that the aggregate size data in Fig.~\ref{fig_df} (b) is collected over a wide range of model parameter values $0.08 \leq n\leq 0.25$, $1/25 \leq \delta_d\leq 1/6$, $0.1\leq F_d\leq 100$, and $0.5 \leq F_T\leq 25$ (a total of $135$ simulations) and still closely collapses on a power law. The only deviation from the power law is near the larger side of the size range which we postulate is due to the morphological transitions. The power law distributions are a common feature (although not always) under the concept of self-organized criticality~\cite{bak59self,bak1988self,gisiger2001scale} in which a system auto-tunes itself towards a critical state without fine-tuning of any control parameter. In our system, although we have found a universal scaling of the size distribution function, we postulate that certain states -- such as the transition to branched phase -- were only possible upon tuning the model parameters {\it e.g.} depletion thickness. In other words, although the system seems to self-organize to produce the size distribution scaling, the system realizes the branched porous states only upon fine-tuning the model parameters.

\section{Confirming sublinear growth using kinetics}
We begin with the discrete form of Smoluchowski's coagulation equation which describes the time evolution of aggregate size distribution $\varphi(s,t)$
\begin{align}\nonumber
	\partial_t \varphi(s,t) = &\frac{1}{2} \sum_{q=1}^{s-1}  K_{s-q,q} \:\varphi(s-q,t) \varphi(q,t) \\
	&-\sum_{q=1}^{\infty} K_{s,q} \:\varphi(s,t) \varphi(q,t),
	\label{eq_smolu_full}
\end{align}
where $s$ or $q$ are the number of monomers in aggregates -- hereon called size of aggregate, and $K_{s,q}$ is the reaction kernel between aggregates of size $s$ and $q$. The first term on the right-hand side accounts for aggregates of size $q$ which react with aggregates of size $s-q$ and attain the size $s$ after reaction. This can happen only if $q<s$ thus the sum is restricted from $q=1$ to $s-1$. The factor half avoids double counting of pairs. The second term accounts for aggregates of size $s$ which react with aggregates of size $q$ and leave the size $s$ after reaction. If we assume that at any given time the distribution $\varphi(s,t)$ remains strictly monodispersed, {\it i.e.} aggregates of only one size are allowed in the system at a given time $t$, then the first term on right-hand side of Eq.~\ref{eq_smolu_full} is zero and the equation reduces to
\begin{align}\nonumber
	\partial_t \varphi(s,t) &= - K_{s,s} \:\varphi(s,t) \varphi(s,t),\\ s\varphi(s,t)&=N,
	\label{eq_smolu_reduced}
\end{align}
with the constraint that the total number of monomers in the system $s\varphi(s,t)$ remain constant.
The reaction kernel $K_{s,s}$ is the product of total collision cross section $\sigma$ between colliding aggregates, the velocities $v$ by which they approach each other, and the probability $p$ that they stick together upon contact, {\it i.e.}, $K_{s,s}\equiv pv\sigma$. 
We construct a model for $K_{s,s}$ as follows. A red blood cell is considered a disk with radius $H$ and thickness $h$ [Fig.~\ref{fig_df} (e, inset)]. Initially, $N$ disk-like free erythrocytes remain homogeneously suspended in the blood plasma. The disks move due to thermal agitation caused by the surrounding liquid molecules and are considered to undergo Brownian motion. The diffusivity of rouleau is $D=k_\mathrm{B} T/6\pi \eta r_H$, where $T$ is the temperature of the surrounding liquid, $k_\mathrm{B}$ is the Boltzmann constant, $\eta$ is the viscosity of the surrounding liquid, and $r_H$ is the hydrodynamic radius of the rouleau. We approximate the rouleau shape as effectively spherical. The hydrodynamic radius $r_H$ is taken equal to the radius of a sphere which will have the same volume as the volume of the rouleau, {\it i.e.}, $(4/3) \pi r_H^3 = \pi H^2 h s$, or $r_H = (3H^2 h s/4)^{1/3}$. The approach speed $v$ between two such rouleaux is estimated using the relation between the long time mean square displacement (MSD), and the diffusivity, {\it i.e.}, $\text{MSD}=2Dt$. This provides that the mean distance traveled by rouleaux in time $t$ is $\sqrt{\text{MSD}}=\sqrt{2Dt}$, and the approach speed is $v=\sqrt{\text{MSD}}/t=\sqrt{2D/t}$ -- a coarse-grained estimate of the Brownian jittering motion. If $t$ is the time traveled between two successive collisions or contacts, which can be estimated by the ratio of the mean separation distance $r$ between disks to the approach speed $v$, then $t={2D/v^2}={2Dt^2/r^2}={2Dt/(rv)}$ or $v=2D/r$. 
The mean separation distance between aggregates itself may have ambiguity in how it is defined. We take the following approach: it is considered that  $r\rightarrow \infty$ as $N^a \:\text{(total no. of aggregates or rouleau)} \rightarrow 1$, $r\rightarrow L/2$ as $N^a \rightarrow 2$, and $r\rightarrow 0$ as $N^a \rightarrow \infty$, defining 
%
%\begin{equation}
$
r=\frac{L}{2}\left(\frac{1}{N^a-1}\right)^{1/3} = \frac{L}{2}\left(\frac{s}{N-s}\right)^{1/3}.
$
%\end{equation}          
%
Here $N=sN^a$ is the total number of disks (monomers) in the system, $L$ is the system size, and $s$ is the number of disks in a single rouleaux. The above relation provides $r\rightarrow \infty$ as $s\rightarrow N$ (meaning approach towards a single aggregate $N^a\rightarrow 1$). Also initially $s=1$, $N\gg1$, and thus $r\sim (1/N)^{1/3}$. Using the above relation for $r$, the approach velocity $v=2D/r$ can be written as  
\begin{equation}
	v=\frac{2D}{\frac{L}{2}\left(\frac{s}{N-s}\right)^{1/3}}
	= \frac{4 k_\mathrm{B} T}{6\pi \eta r_H L} \left(\frac{N-s}{s}\right)^{1/3}.
\end{equation}       
where $r_H = (3H^2 h s/4)^{1/3}$. The collision cross section for rouleaux with $s$ number of disks is
\begin{equation}
	\sigma = \pi r_H^2 = \pi \left( \frac{3H^2 h s}{4} \right)^{2/3}.
\end{equation}     
The probability that a collision leads to sticking is taken equal to the ratio of the surface area on a rouleaux which permits sticking, to the total surface area of the rouleaux, {\it i.e.}     
\begin{equation}
	p=\frac{2 \pi H^2}{2 \pi H^2 + 2 \pi H h s} = \frac{1}{1+(h/H) s}.
	\label{eq_p}
\end{equation}
Finally, using $v,\sigma$ and $p$, we construct the reaction kernel $K_{s,s}$ as
\begin{equation}
	K_{s,s}=p v \sigma=C \left(\frac{N-s}{s}\right)^{1/3} \frac{s^{1/3}}{1+(h/H)s},
	\label{eq_K}
\end{equation}
where 
\begin{equation}
	C=\pi \frac{ 4 k_\mathrm{B} T }{6 \pi \eta L}  \left(\frac{3 H^2 h}{4}\right)^{1/3}.
\end{equation}
The fact from Eq.~\ref{eq_smolu_reduced} that the total number of monomers $s(t) \varphi(s,t)=N$ is a constant, provides
\begin{equation}
	\frac{\partial (s \varphi)}{\partial t}= s\frac{\partial \varphi}{\partial t} +\varphi \frac{\partial s}{\partial t}= 0,
\end{equation}
or using $\partial_t \varphi = - K_{s,s} \:\varphi^2$ from Eq.~\ref{eq_smolu_reduced}, we write
\begin{equation}
	\frac{\partial s}{\partial t}= -\frac{s}{\varphi}\frac{\partial \varphi}{\partial t}={s}K_{s,s}\varphi=sK_{s,s}(N/s)=NK_{s,s}.
	\label{eq_dndt}
\end{equation}
For aggregation of Brownian spheres with sticking probability $p=1$, the reaction kernel reduces to
\begin{equation}
	K_{s,s}= \frac{4k_\mathrm{B}TH}{6\eta L}(N-s)^{1/3},
	\label{eq_K_spheres}
\end{equation}
which can be used to solve Eq.~\ref{eq_dndt} for the case of spherical particles. Here $H$ is the radii of the spheres. Using this kernel, the solution of Eq.~\ref{eq_dndt} reads
\begin{equation}
	s(t)=N-\left[ (N-1)^{2/3} - \frac{4k_\mathrm{B}TH N}{9\eta L}t \right]^{3/2}.
	\label{eq_nt_spheres}
\end{equation}   
The solution for Brownian disks, with aspect ratio dependent sticking probability [Eq.~\ref{eq_p}] and kernel [Eq.~\ref{eq_K}], is more involved, and in an implicit form it reads
\begin{align}\nonumber
	t\frac{4k_\mathrm{B}THN}{6\eta L}
	\left(\frac{3}{4}\right)^{1/3}
	\left(\frac{h}{H}\right)^{1/3}
	=&-\frac{3}{10}[N-s(t)]^{2/3}
	\left[
	2s(t)\frac{h}{H}
	+3N\frac{h}{H}
	+5
	\right]
	\\
	&+\frac{3}{10}[N-1]^{2/3}
	\left[
	2\frac{h}{H}
	+3N\frac{h}{H}
	+5
	\right].
	\label{eq_nt_disks}
\end{align} 
For a particular set of conditions: $k_\mathrm{B}T,N,L,\eta$ and $H$, solutions~\ref{eq_nt_spheres} and~\ref{eq_nt_disks} are compared in Fig.~\ref{fig_df} (e).
It is clear that mimicking the shape of the cells using disks results in sublinear growth. In the case of spherical shape, it is the symmetry that the sticking probability $p=1$ for all contacts which leads to relatively increased and nearly linear growth rate. Upon comparison with seminal experiments performed by~\citet{kernick1973experiments} [Fig.~\ref{fig_df} (g)], it is found that the Smoluchowski's equation, even for disk-shaped cells, over-predicts the growth exponent. It might be due to the fact that we have considered three-dimensional motion of disks in the derivation. However comparing the same disk model with experiments from~\citet{barshtein2000kinetics} [Fig.~\ref{fig_df} (f)], the growth matches a time regime where the average rouleau size grows as a sublinear power law. 
Additionally, Eq.~\ref{eq_nt_spheres} and \ref{eq_nt_disks} provide growth crossover time scales for spheres and disks respectively  
\begin{align}
	\tau_\mathrm{spheres} = \frac{\eta L}{k_\mathrm{B} T H N},\: \tau_\mathrm{disks} = \tau_\mathrm{spheres} \left(\frac{H}{h}\right)^{1/3},
\end{align} 
indicating that the crossover time to the power law growth in the case of disks is higher than in the case of spheres, by a factor of $(H/h)^{1/3}$. Thus compared to spheres, the onset of power law growth for disk-shaped cells is delayed and the growth exponent is relatively lower.
For comparing the 2D simulations to the Smoluchowski predictions, we can set $\bar{R}(t)\sim{s(t)^{1/D_f}}$. Thus if $D_f=2$ and the Smoluchowski prediction from Eq.~\ref{eq_nt_disks} is nearly ${s}(t)\sim{t^{0.7}}$, which basically says that $\bar{R}(t)\sim{t^{0.35}}$ which is quite sublinear. In practice, the sticking probability, approach velocities, and collision cross section might be more involved, thus leading to different values of the growth exponent. However, it is safe to postulate that our analysis, and data from simulations, puts an upper limit on the growth exponent and suggests that it has to be sublinear, which is consistent with the past experiments.
\begin{figure*}[t!]
	\includegraphics[width=1.0\linewidth]{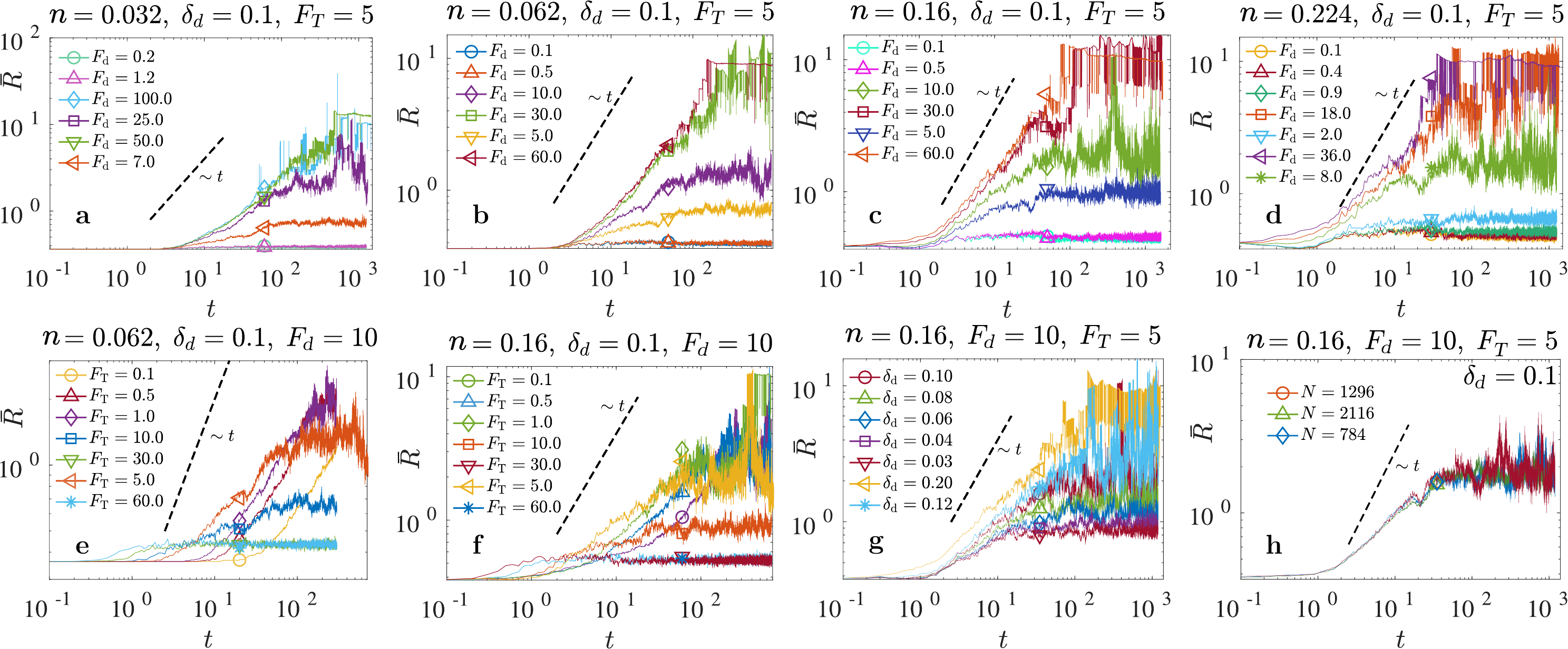}	
	\caption{Mean radius of gyration $\bar{R}$ of aggregates with time under varying model parameters: $F_d$, $F_T$, $n$, $\delta_d$ and $N$, when an asymmetric alignment factor Eq.~\ref{eq_orientation_factor_2} is used in the depletion force Eq.~\ref{eq_fd_1}.  
		(e, f) Effect of varying thermal force strength $F_T$, (g) varying depletion thickness $\delta_d$, and (h) varying total number of cells $N$. 
		The growth rate remains consistently sublinear except at relatively high $n$ and $F_d$ [panel (d)], which lead to percolation and giant single clusters under strong depletion force strength $F_d>20$ combined with relatively high number densities $n>0.2$.
		However, upon making the alignment factor asymmetric as described in Eq.~\ref{eq_orientation_factor_2}, we do not observe sharply branched aggregates.}
	\label{fig_growth_with_time_2}
\end{figure*}

\section{Conclusions}
We have proposed that systematically designed alignment-dependent attractive interactions in a system of dimers can precisely simulate depletion-mediated branched and porous structures observed in the microscopic images of aggregated erythrocytes or RBCs. The model consistently predicts experimental observations and kinetic model predictions of sublinear or nearly linear growth of mean erythrocyte aggregate size. We have summarized various configurational states realized by the system in terms of ($i$) the ratio of depletion force strength to thermal force strength $F_d/F_T$, ($ii$) number density $n$, and ($iii$) depletion thickness $\delta_d$. The maximum of average linear size follows a scaling ${\bar{\ell}_\mathrm{max}} \sim \sqrt{n\delta_d {F_d}/{F_T}}$ for intermediate values of $n\delta_d {F_d}/{F_T}$. If we collect an ensemble of aggregate radii of gyration $R$ and plot the size distribution function $f(R)$, we find a universal power law behavior $f(R)\sim (R)^{-3.3}$ irrespective of the values of the model parameters. Although the system seems to self-organize to produce the universal size distribution scaling, the system realizes the branched porous states only upon fine-tuning the model parameters. Upon lowering the depletion thickness, the average area-to-mass ratio of aggregates almost diverges as we increase $n\delta_dF_d/F_T$ clearly marking the transition to the branched phase.
Our results are testable using techniques such as flow cytometry, image processing, or optical techniques such as light transmission through a blood sample. 
In general, we believe that our work will generate interest in under-explored field of branched aggregate structures, not only in relation to RBCs, but also in other areas of soft condensed matter physics such as polymers.
{\section*{Appendix I: Numerical methods and scales}}
{Once the total force on a sub-particle $p\in[1,2]$ belonging to a dimer cell $i\in[1,N]$ is known [Fig.~\ref{fig_model} (a-d)], we integrate the following Newton's equation of motion for that sub-particle
	\begin{align}
		m_{i_p}\frac{d\boldsymbol{v}_{i_p}}{dt} 
		=\boldsymbol{F}_{i_p}^\mathrm{total}
		=\sum_{j=1,j\neq i}^{N}\sum_{q=1}^{2}\left[ \boldsymbol{F}_{i_pj_q}^\mathrm{d}+\boldsymbol{F}_{i_pj_q}^\mathrm{contact} \right]
		+\boldsymbol{F}_{i_p}^\mathrm{thermal}+\boldsymbol{F}_{i_pi_q}^\mathrm{internal},
	\end{align}
	where $\boldsymbol{F}_{i_pj_q}^\mathrm{d},\:\boldsymbol{F}_{i_pj_q}^\mathrm{contact}$ and 
	$\boldsymbol{F}_{i_p}^\mathrm{thermal}$ are expressed in Eq~\ref{eq_fd_1}, ~\ref{eq_contact}, and ~\ref{eq_ft} respectively, while $\boldsymbol{F}_{i_pi_q}^\mathrm{internal}$ is the internal linear spring-dashpot force holding together the dimer sub-particles $i_p$ and $i_q$. Note that the expressions of $\boldsymbol{F}_{i_pj_q}^\mathrm{d}$ and $\boldsymbol{F}_{i_pj_q}^\mathrm{contact}$ [Eq.~\ref{eq_fd_1}, \ref{eq_contact}] involve Heaviside step functions and thus the sum over these two forces by definition exclude particles out of respective interaction cut-offs.
	We utilize the velocity-verlet scheme for integrating the system in time. The scheme proceeds as follows. The velocity of a dimer sub-particle is first calculated at half-time step using force on the particle at time $t$, {\it i.e.} 
	\begin{align}
		\boldsymbol{v}_{i_p}(t+\Delta t/2)=\boldsymbol{v}_{i_p}(t)+\frac{1}{2}\Delta t\:\boldsymbol{F}_{i_p}^\mathrm{total}(t)/m_{i_p}.
	\end{align}
	The position of the dimer sub-particle is then updated using this velocity at the half-time step {\it i.e.} 
	\begin{align}
		\boldsymbol{x}_{i_p}(t+\Delta t)=\boldsymbol{x}_{i_p}(t)+\Delta t\:\boldsymbol{v}_{i_p}(t+\Delta t/2).
	\end{align}
	Periodic boundary conditions are applied to all the particles at this point of the algorithm. The force $\boldsymbol{F}_{i_p}^\mathrm{total}(t+\Delta t)$ is then computed at this updated position $\boldsymbol{x}_{i_p}(t+\Delta t)$, and finally the velocity is corrected using 
	\begin{align}
		\boldsymbol{v}_{i_p}(t+\Delta t)=\boldsymbol{v}_{i_p}(t+\Delta t/2)+\frac{1}{2}\Delta t\:\boldsymbol{F}_{i_p}^\mathrm{total}(t+\Delta t)/m_{i_p}.
	\end{align}
	The simulation moves to the next time step and the procedure is repeated till desired time. We implement the entire algorithm in C++ and utilize its object-oriented features for efficiency, while the data and cluster analysis are carried out in MATLAB.}
{The dimers are initialized at $t=0$ with uniformly random positions, uniformly random orientations, and zero velocities. Note that although individual sub-particles in a dimer do not have orientations, the dimer as a whole has an orientation given by the line joining the centers of the two sub-particles in the dimer. We have neglected the effects arising due to the viscosity of the surrounding medium, or effectively we treat the system as if the thermal forces on dimer sub-particles are the resultant of the forces due to molecular collisions and the drag due to viscosity. Explicit treatment of viscous forces and their effect on the aggregate structures is left for future exploration.}
The typical size and thickness of an RBC is approximately $7-8$ $\mu$m and $2.5$ $\mu$m respectively~\cite{kinnunen2011effect}. The mass of an RBC has been reported to be $\approx 27$ pg (picograms)~\cite{phillips2012measurement} and aggregation force between two RBCs using optical tweezers has been reported to be $\approx 1-20$ pN~\cite{lee2016optical}. Following these observations if we rescale and measure the mass, length, and force in units of $m_o=27\times 10^{-12}$ g, $d_o=2.5\times 10^{-6}$ m, $F_o=1\times10^{-12}$ N respectively, then naturally the unit of time comes out to be $t_o=\sqrt{m_od_o/F_o}=0.0082$ s. Thus our simulation units of time $t=1200$ correspond to $1200\times t_o \approx 10$ s. Under this rescaling, the elastic and damping constants are $k=\tilde{k} d_o^{3/2}/F_o$ and $\gamma=\tilde{\gamma}d_o/(t_oF_o)$ respectively where $\tilde{k}$ and $\tilde{\gamma}$ are elastic and damping constants in units of N/m$^{3/2}$ and Ns/m respectively. We have fixed $k=1000$ and $\gamma=20$ in the simulations which implies that the RBCs are assumed to have material properties $\tilde{k}=kF_o/d_o^{3/2}=0.2530$ N/m$^{3/2}$ and $\tilde{\gamma}=\gamma t_o F_o/d_o = 6.5\times 10^{-8}$ Ns/m. These material properties can be further related to modulus of elasticity, Poisson's ratio, and material viscous properties, however, studies on characterization of these properties for RBCs are scarce. Nevertheless, to a limit, our model can be well adjusted to accommodate appropriate RBC material properties.
\section*{Appendix II: Asymmetric alignment factor}
The growth curves $\bar{R}(t)$ are also computed after modifying the alignment factor of Eq.~\ref{eq_orientation_factor}, for instance, using only the first part 
\begin{align}
	O_{i_{1}j_{1}}&=(1-|\boldsymbol{e}_{i_{1}j_{1}}\cdot\boldsymbol{e}_{i_{1}i_{2}}|)
	\equiv O_{i_{1}j_{1}}^{(i)}.
	\label{eq_orientation_factor_2}
\end{align}
This form of the alignment factor makes the pairwise depletion forces asymmetric. It served as another test to see if the growth rates surpass the linear limit if we relax the alignment factor. We see that under this modification, the growth rates still largely remain sub-linear except in cases where $n$ and $F_d$ were high [Fig.~\ref{fig_growth_with_time_2} (d)]. However if make the pairwise depletion forces asymmetric as described above in Eq.~\ref{eq_orientation_factor_2}, we do not observe sharply branched aggregates. The transition to the branched phase with sharp nodes is easily simulated under symmetric depletion forces Eq.~\ref{eq_orientation_factor} which also ensures that Newton's third law is obeyed.
If we change the total number of particles present in the system, under both models of the alignment factor Eq.~\ref{eq_orientation_factor} and Eq.~\ref{eq_orientation_factor_2}, we see that the growth calculations are negligibly affected as long as $n$ remains the same [Fig.~\ref{fig_growth_with_time_2} (h)]. If we relax the anisotropy of interactions, the growth rate still remains sublinear but with thicker branch nodes in the branched phase. In addition, the effect of thermal force $F_T$ on the growth rate is depicted in [Fig.~\ref{fig_growth_with_time_2} (e, f)] where the rate first increases with increasing $F_T$ (indicating increase in approach velocities) and thereafter it decreases with increasing $F_T$ (indicating breakage of aggregates due to thermal agitation) consistent with past experiments~\cite{kernick1973experiments}.

\section*{Acknowledgments}
CS is thankful for the financial support by the Department of Science and Technology, India under the INSPIRE Faculty Fellowship Award.

%%%END OF MAIN TEXT%%%

%apsrev4-2.bst 2019-01-14 (MD) hand-edited version of apsrev4-1.bst
%Control: key (0)
%Control: author (8) initials jnrlst
%Control: editor formatted (1) identically to author
%Control: production of article title (0) allowed
%Control: page (0) single
%Control: year (1) truncated
%Control: production of eprint (0) enabled
%

%%%REFERENCES%%%
%\bibliography{references} 

\begin{thebibliography}{51}%
	\makeatletter
	\providecommand \@ifxundefined [1]{%
		\@ifx{#1\undefined}
	}%
	\providecommand \@ifnum [1]{%
		\ifnum #1\expandafter \@firstoftwo
		\else \expandafter \@secondoftwo
		\fi
	}%
	\providecommand \@ifx [1]{%
		\ifx #1\expandafter \@firstoftwo
		\else \expandafter \@secondoftwo
		\fi
	}%
	\providecommand \natexlab [1]{#1}%
	\providecommand \enquote  [1]{``#1''}%
	\providecommand \bibnamefont  [1]{#1}%
	\providecommand \bibfnamefont [1]{#1}%
	\providecommand \citenamefont [1]{#1}%
	\providecommand \href@noop [0]{\@secondoftwo}%
	\providecommand \href [0]{\begingroup \@sanitize@url \@href}%
	\providecommand \@href[1]{\@@startlink{#1}\@@href}%
	\providecommand \@@href[1]{\endgroup#1\@@endlink}%
	\providecommand \@sanitize@url [0]{\catcode `\\12\catcode `\$12\catcode
		`\&12\catcode `\#12\catcode `\^12\catcode `\_12\catcode `\%12\relax}%
	\providecommand \@@startlink[1]{}%
	\providecommand \@@endlink[0]{}%
	\providecommand \url  [0]{\begingroup\@sanitize@url \@url }%
	\providecommand \@url [1]{\endgroup\@href {#1}{\urlprefix }}%
	\providecommand \urlprefix  [0]{URL }%
	\providecommand \Eprint [0]{\href }%
	\providecommand \doibase [0]{https://doi.org/}%
	\providecommand \selectlanguage [0]{\@gobble}%
	\providecommand \bibinfo  [0]{\@secondoftwo}%
	\providecommand \bibfield  [0]{\@secondoftwo}%
	\providecommand \translation [1]{[#1]}%
	\providecommand \BibitemOpen [0]{}%
	\providecommand \bibitemStop [0]{}%
	\providecommand \bibitemNoStop [0]{.\EOS\space}%
	\providecommand \EOS [0]{\spacefactor3000\relax}%
	\providecommand \BibitemShut  [1]{\csname bibitem#1\endcsname}%
	\let\auto@bib@innerbib\@empty
	%</preamble>
	\bibitem [{\citenamefont {Litvinov}\ and\ \citenamefont
		{Weisel}(2017)}]{litvinov2017role}%
	\BibitemOpen
	\bibfield  {author} {\bibinfo {author} {\bibfnamefont {R.~I.}\ \bibnamefont
			{Litvinov}}\ and\ \bibinfo {author} {\bibfnamefont {J.~W.}\ \bibnamefont
			{Weisel}},\ }\bibfield  {title} {\bibinfo {title} {Role of red blood cells in
			haemostasis and thrombosis},\ }\href@noop {} {\bibfield  {journal} {\bibinfo
			{journal} {ISBT science series}\ }\textbf {\bibinfo {volume} {12}},\ \bibinfo
		{pages} {176} (\bibinfo {year} {2017})}\BibitemShut {NoStop}%
	\bibitem [{\citenamefont {Satoh}\ \emph {et~al.}(1984)\citenamefont {Satoh},
		\citenamefont {Imaizumi}, \citenamefont {Bessho},\ and\ \citenamefont
		{Shiga}}]{satoh1984increased}%
	\BibitemOpen
	\bibfield  {author} {\bibinfo {author} {\bibfnamefont {M.}~\bibnamefont
			{Satoh}}, \bibinfo {author} {\bibfnamefont {K.}~\bibnamefont {Imaizumi}},
		\bibinfo {author} {\bibfnamefont {T.}~\bibnamefont {Bessho}},\ and\ \bibinfo
		{author} {\bibfnamefont {T.}~\bibnamefont {Shiga}},\ }\bibfield  {title}
	{\bibinfo {title} {Increased erythrocyte aggregation in diabetes mellitus and
			its relationship to glycosylated haemoglobin and retinopathy},\ }\href@noop
	{} {\bibfield  {journal} {\bibinfo  {journal} {Diabetologia}\ }\textbf
		{\bibinfo {volume} {27}},\ \bibinfo {pages} {517} (\bibinfo {year}
		{1984})}\BibitemShut {NoStop}%
	\bibitem [{\citenamefont {Sheremet’ev}\ \emph {et~al.}(2019)\citenamefont
		{Sheremet’ev}, \citenamefont {Popovicheva}, \citenamefont {Rogozin},\ and\
		\citenamefont {Levin}}]{sheremet2019red}%
	\BibitemOpen
	\bibfield  {author} {\bibinfo {author} {\bibfnamefont {Y.~A.}\ \bibnamefont
			{Sheremet’ev}}, \bibinfo {author} {\bibfnamefont {A.~N.}\ \bibnamefont
			{Popovicheva}}, \bibinfo {author} {\bibfnamefont {M.~M.}\ \bibnamefont
			{Rogozin}},\ and\ \bibinfo {author} {\bibfnamefont {G.~Y.}\ \bibnamefont
			{Levin}},\ }\bibfield  {title} {\bibinfo {title} {Red blood cell aggregation,
			disaggregation and aggregate morphology in autologous plasma and serum in
			diabetic foot disease},\ }\href@noop {} {\bibfield  {journal} {\bibinfo
			{journal} {Clinical Hemorheology and Microcirculation}\ }\textbf {\bibinfo
			{volume} {72}},\ \bibinfo {pages} {221} (\bibinfo {year} {2019})}\BibitemShut
	{NoStop}%
	\bibitem [{\citenamefont {Almog}\ \emph {et~al.}(2005)\citenamefont {Almog},
		\citenamefont {Gamzu}, \citenamefont {Almog}, \citenamefont {Lessing},
		\citenamefont {Shapira}, \citenamefont {Berliner}, \citenamefont {Pauzner},
		\citenamefont {Maslovitz},\ and\ \citenamefont {Levin}}]{almog2005enhanced}%
	\BibitemOpen
	\bibfield  {author} {\bibinfo {author} {\bibfnamefont {B.}~\bibnamefont
			{Almog}}, \bibinfo {author} {\bibfnamefont {R.}~\bibnamefont {Gamzu}},
		\bibinfo {author} {\bibfnamefont {R.}~\bibnamefont {Almog}}, \bibinfo
		{author} {\bibfnamefont {J.~B.}\ \bibnamefont {Lessing}}, \bibinfo {author}
		{\bibfnamefont {I.}~\bibnamefont {Shapira}}, \bibinfo {author} {\bibfnamefont
			{S.}~\bibnamefont {Berliner}}, \bibinfo {author} {\bibfnamefont
			{D.}~\bibnamefont {Pauzner}}, \bibinfo {author} {\bibfnamefont
			{S.}~\bibnamefont {Maslovitz}},\ and\ \bibinfo {author} {\bibfnamefont
			{I.}~\bibnamefont {Levin}},\ }\bibfield  {title} {\bibinfo {title} {Enhanced
			erythrocyte aggregation in clinically diagnosed pelvic inflammatory
			disease},\ }\href@noop {} {\bibfield  {journal} {\bibinfo  {journal}
			{Sexually transmitted diseases}\ }\textbf {\bibinfo {volume} {32}},\ \bibinfo
		{pages} {484} (\bibinfo {year} {2005})}\BibitemShut {NoStop}%
	\bibitem [{\citenamefont {Peled}\ \emph {et~al.}(2008)\citenamefont {Peled},
		\citenamefont {Kassirer}, \citenamefont {Kramer}, \citenamefont {Rogowski},
		\citenamefont {Shlomi}, \citenamefont {Fox}, \citenamefont {Berliner},\ and\
		\citenamefont {Shitrit}}]{peled2008increased}%
	\BibitemOpen
	\bibfield  {author} {\bibinfo {author} {\bibfnamefont {N.}~\bibnamefont
			{Peled}}, \bibinfo {author} {\bibfnamefont {M.}~\bibnamefont {Kassirer}},
		\bibinfo {author} {\bibfnamefont {M.~R.}\ \bibnamefont {Kramer}}, \bibinfo
		{author} {\bibfnamefont {O.}~\bibnamefont {Rogowski}}, \bibinfo {author}
		{\bibfnamefont {D.}~\bibnamefont {Shlomi}}, \bibinfo {author} {\bibfnamefont
			{B.}~\bibnamefont {Fox}}, \bibinfo {author} {\bibfnamefont {A.~S.}\
			\bibnamefont {Berliner}},\ and\ \bibinfo {author} {\bibfnamefont
			{D.}~\bibnamefont {Shitrit}},\ }\bibfield  {title} {\bibinfo {title}
		{Increased erythrocyte adhesiveness and aggregation in obstructive sleep
			apnea syndrome},\ }\href@noop {} {\bibfield  {journal} {\bibinfo  {journal}
			{Thrombosis research}\ }\textbf {\bibinfo {volume} {121}},\ \bibinfo {pages}
		{631} (\bibinfo {year} {2008})}\BibitemShut {NoStop}%
	\bibitem [{\citenamefont {Clark}\ \emph {et~al.}(2018)\citenamefont {Clark},
		\citenamefont {Fenner}, \citenamefont {Sasson}, \citenamefont {McClain},
		\citenamefont {Singer},\ and\ \citenamefont {Tonnesen}}]{clark2018blood}%
	\BibitemOpen
	\bibfield  {author} {\bibinfo {author} {\bibfnamefont {R.~A.}\ \bibnamefont
			{Clark}}, \bibinfo {author} {\bibfnamefont {J.}~\bibnamefont {Fenner}},
		\bibinfo {author} {\bibfnamefont {A.}~\bibnamefont {Sasson}}, \bibinfo
		{author} {\bibfnamefont {S.~A.}\ \bibnamefont {McClain}}, \bibinfo {author}
		{\bibfnamefont {A.~J.}\ \bibnamefont {Singer}},\ and\ \bibinfo {author}
		{\bibfnamefont {M.~G.}\ \bibnamefont {Tonnesen}},\ }\bibfield  {title}
	{\bibinfo {title} {Blood vessel occlusion with erythrocyte aggregates causes
			burn injury progression—microvasculature dilation as a possible therapy},\
	}\href@noop {} {\bibfield  {journal} {\bibinfo  {journal} {Experimental
				dermatology}\ }\textbf {\bibinfo {volume} {27}},\ \bibinfo {pages} {625}
		(\bibinfo {year} {2018})}\BibitemShut {NoStop}%
	\bibitem [{\citenamefont {Baskurt}\ \emph {et~al.}(2011)\citenamefont
		{Baskurt}, \citenamefont {Neu},\ and\ \citenamefont
		{Meiselman}}]{baskurt2011red}%
	\BibitemOpen
	\bibfield  {author} {\bibinfo {author} {\bibfnamefont {O.}~\bibnamefont
			{Baskurt}}, \bibinfo {author} {\bibfnamefont {B.}~\bibnamefont {Neu}},\ and\
		\bibinfo {author} {\bibfnamefont {H.~J.}\ \bibnamefont {Meiselman}},\
	}\bibfield  {title} {\bibinfo {title} {Red blood cell aggregation},\
	}\href@noop {} {\  (\bibinfo {year} {2011})}\BibitemShut {NoStop}%
	\bibitem [{\citenamefont {Chien}\ and\ \citenamefont
		{Jan}(1973)}]{chien1973ultrastructural}%
	\BibitemOpen
	\bibfield  {author} {\bibinfo {author} {\bibfnamefont {S.}~\bibnamefont
			{Chien}}\ and\ \bibinfo {author} {\bibfnamefont {K.-m.}\ \bibnamefont
			{Jan}},\ }\bibfield  {title} {\bibinfo {title} {Ultrastructural basis of the
			mechanism of rouleaux formation},\ }\href@noop {} {\bibfield  {journal}
		{\bibinfo  {journal} {Microvascular research}\ }\textbf {\bibinfo {volume}
			{5}},\ \bibinfo {pages} {155} (\bibinfo {year} {1973})}\BibitemShut {NoStop}%
	\bibitem [{\citenamefont {Wagner}\ \emph {et~al.}(2013)\citenamefont {Wagner},
		\citenamefont {Steffen},\ and\ \citenamefont
		{Svetina}}]{wagner2013aggregation}%
	\BibitemOpen
	\bibfield  {author} {\bibinfo {author} {\bibfnamefont {C.}~\bibnamefont
			{Wagner}}, \bibinfo {author} {\bibfnamefont {P.}~\bibnamefont {Steffen}},\
		and\ \bibinfo {author} {\bibfnamefont {S.}~\bibnamefont {Svetina}},\
	}\bibfield  {title} {\bibinfo {title} {Aggregation of red blood cells: from
			rouleaux to clot formation},\ }\href@noop {} {\bibfield  {journal} {\bibinfo
			{journal} {Comptes Rendus Physique}\ }\textbf {\bibinfo {volume} {14}},\
		\bibinfo {pages} {459} (\bibinfo {year} {2013})}\BibitemShut {NoStop}%
	\bibitem [{\citenamefont {Steffen}\ \emph {et~al.}(2013)\citenamefont
		{Steffen}, \citenamefont {Verdier},\ and\ \citenamefont
		{Wagner}}]{steffen2013quantification}%
	\BibitemOpen
	\bibfield  {author} {\bibinfo {author} {\bibfnamefont {P.}~\bibnamefont
			{Steffen}}, \bibinfo {author} {\bibfnamefont {C.}~\bibnamefont {Verdier}},\
		and\ \bibinfo {author} {\bibfnamefont {C.}~\bibnamefont {Wagner}},\
	}\bibfield  {title} {\bibinfo {title} {Quantification of depletion-induced
			adhesion of red blood cells},\ }\href@noop {} {\bibfield  {journal} {\bibinfo
			{journal} {Physical review letters}\ }\textbf {\bibinfo {volume} {110}},\
		\bibinfo {pages} {018102} (\bibinfo {year} {2013})}\BibitemShut {NoStop}%
	\bibitem [{\citenamefont {Flormann}\ \emph {et~al.}(2017)\citenamefont
		{Flormann}, \citenamefont {Aouane}, \citenamefont {Kaestner}, \citenamefont
		{Ruloff}, \citenamefont {Misbah}, \citenamefont {Podgorski},\ and\
		\citenamefont {Wagner}}]{flormann2017buckling}%
	\BibitemOpen
	\bibfield  {author} {\bibinfo {author} {\bibfnamefont {D.}~\bibnamefont
			{Flormann}}, \bibinfo {author} {\bibfnamefont {O.}~\bibnamefont {Aouane}},
		\bibinfo {author} {\bibfnamefont {L.}~\bibnamefont {Kaestner}}, \bibinfo
		{author} {\bibfnamefont {C.}~\bibnamefont {Ruloff}}, \bibinfo {author}
		{\bibfnamefont {C.}~\bibnamefont {Misbah}}, \bibinfo {author} {\bibfnamefont
			{T.}~\bibnamefont {Podgorski}},\ and\ \bibinfo {author} {\bibfnamefont
			{C.}~\bibnamefont {Wagner}},\ }\bibfield  {title} {\bibinfo {title} {The
			buckling instability of aggregating red blood cells},\ }\href@noop {}
	{\bibfield  {journal} {\bibinfo  {journal} {Scientific reports}\ }\textbf
		{\bibinfo {volume} {7}},\ \bibinfo {pages} {1} (\bibinfo {year}
		{2017})}\BibitemShut {NoStop}%
	\bibitem [{\citenamefont {Flormann}(2017)}]{flormann2017physical}%
	\BibitemOpen
	\bibfield  {author} {\bibinfo {author} {\bibfnamefont {D.~A.~D.}\
			\bibnamefont {Flormann}},\ }\emph {\bibinfo {title} {Physical charaterization
			of red blood cell aggregation}},\ \href@noop {} {Ph.D. thesis},\ \bibinfo
	{school} {Universit{\"a}t des Saarlandes} (\bibinfo {year}
	{2017})\BibitemShut {NoStop}%
	\bibitem [{\citenamefont {Lanotte}\ \emph {et~al.}(2016)\citenamefont
		{Lanotte}, \citenamefont {Mauer}, \citenamefont {Mendez}, \citenamefont
		{Fedosov}, \citenamefont {Fromental}, \citenamefont {Claveria}, \citenamefont
		{Nicoud}, \citenamefont {Gompper},\ and\ \citenamefont
		{Abkarian}}]{lanotte2016red}%
	\BibitemOpen
	\bibfield  {author} {\bibinfo {author} {\bibfnamefont {L.}~\bibnamefont
			{Lanotte}}, \bibinfo {author} {\bibfnamefont {J.}~\bibnamefont {Mauer}},
		\bibinfo {author} {\bibfnamefont {S.}~\bibnamefont {Mendez}}, \bibinfo
		{author} {\bibfnamefont {D.~A.}\ \bibnamefont {Fedosov}}, \bibinfo {author}
		{\bibfnamefont {J.-M.}\ \bibnamefont {Fromental}}, \bibinfo {author}
		{\bibfnamefont {V.}~\bibnamefont {Claveria}}, \bibinfo {author}
		{\bibfnamefont {F.}~\bibnamefont {Nicoud}}, \bibinfo {author} {\bibfnamefont
			{G.}~\bibnamefont {Gompper}},\ and\ \bibinfo {author} {\bibfnamefont
			{M.}~\bibnamefont {Abkarian}},\ }\bibfield  {title} {\bibinfo {title} {Red
			cells’ dynamic morphologies govern blood shear thinning under
			microcirculatory flow conditions},\ }\href@noop {} {\bibfield  {journal}
		{\bibinfo  {journal} {Proceedings of the National Academy of Sciences}\
		}\textbf {\bibinfo {volume} {113}},\ \bibinfo {pages} {13289} (\bibinfo
		{year} {2016})}\BibitemShut {NoStop}%
	\bibitem [{\citenamefont {Barshtein}\ \emph {et~al.}(2000)\citenamefont
		{Barshtein}, \citenamefont {Wajnblum},\ and\ \citenamefont
		{Yedgar}}]{barshtein2000kinetics}%
	\BibitemOpen
	\bibfield  {author} {\bibinfo {author} {\bibfnamefont {G.}~\bibnamefont
			{Barshtein}}, \bibinfo {author} {\bibfnamefont {D.}~\bibnamefont
			{Wajnblum}},\ and\ \bibinfo {author} {\bibfnamefont {S.}~\bibnamefont
			{Yedgar}},\ }\bibfield  {title} {\bibinfo {title} {Kinetics of linear
			rouleaux formation studied by visual monitoring of red cell dynamic
			organization},\ }\href@noop {} {\bibfield  {journal} {\bibinfo  {journal}
			{Biophysical journal}\ }\textbf {\bibinfo {volume} {78}},\ \bibinfo {pages}
		{2470} (\bibinfo {year} {2000})}\BibitemShut {NoStop}%
	\bibitem [{\citenamefont {Sewchand}\ \emph {et~al.}(1982)\citenamefont
		{Sewchand}, \citenamefont {Rowlands},\ and\ \citenamefont
		{Lovlin}}]{sewchand1982resistance}%
	\BibitemOpen
	\bibfield  {author} {\bibinfo {author} {\bibfnamefont {L.}~\bibnamefont
			{Sewchand}}, \bibinfo {author} {\bibfnamefont {S.}~\bibnamefont {Rowlands}},\
		and\ \bibinfo {author} {\bibfnamefont {R.}~\bibnamefont {Lovlin}},\
	}\bibfield  {title} {\bibinfo {title} {Resistance to the brownian movement of
			red blood cells on flat horizontal surfaces},\ }\href@noop {} {\bibfield
		{journal} {\bibinfo  {journal} {Cell biophysics}\ }\textbf {\bibinfo {volume}
			{4}},\ \bibinfo {pages} {41} (\bibinfo {year} {1982})}\BibitemShut {NoStop}%
	\bibitem [{\citenamefont {Singh}(2021)}]{singh2021guided}%
	\BibitemOpen
	\bibfield  {author} {\bibinfo {author} {\bibfnamefont {C.}~\bibnamefont
			{Singh}},\ }\bibfield  {title} {\bibinfo {title} {Guided run-and-tumble
			active particles: wall accumulation and preferential deposition},\
	}\href@noop {} {\bibfield  {journal} {\bibinfo  {journal} {Soft Matter}\
		}\textbf {\bibinfo {volume} {17}},\ \bibinfo {pages} {8858} (\bibinfo {year}
		{2021})}\BibitemShut {NoStop}%
	\bibitem [{\citenamefont {Ponder}(1925)}]{ponder1925sedimentation}%
	\BibitemOpen
	\bibfield  {author} {\bibinfo {author} {\bibfnamefont {E.}~\bibnamefont
			{Ponder}},\ }\bibfield  {title} {\bibinfo {title} {On sedimentation and
			rouleaux formation-i},\ }\href@noop {} {\bibfield  {journal} {\bibinfo
			{journal} {Quarterly Journal of Experimental Physiology: Translation and
				Integration}\ }\textbf {\bibinfo {volume} {15}},\ \bibinfo {pages} {235}
		(\bibinfo {year} {1925})}\BibitemShut {NoStop}%
	\bibitem [{\citenamefont {Ponder}(1926)}]{ponder1926sedimentation}%
	\BibitemOpen
	\bibfield  {author} {\bibinfo {author} {\bibfnamefont {E.}~\bibnamefont
			{Ponder}},\ }\bibfield  {title} {\bibinfo {title} {On sedimentation and
			rouleaux formation—ii},\ }\href@noop {} {\bibfield  {journal} {\bibinfo
			{journal} {Quarterly Journal of Experimental Physiology: Translation and
				Integration}\ }\textbf {\bibinfo {volume} {16}},\ \bibinfo {pages} {173}
		(\bibinfo {year} {1926})}\BibitemShut {NoStop}%
	\bibitem [{\citenamefont {Ponder}(1932)}]{ponder1932sedimentation}%
	\BibitemOpen
	\bibfield  {author} {\bibinfo {author} {\bibfnamefont {E.}~\bibnamefont
			{Ponder}},\ }\bibfield  {title} {\bibinfo {title} {On sedimentation and
			rouleaux formation.—iii. the sedimentation of spherical erythrocytes},\
	}\href@noop {} {\bibfield  {journal} {\bibinfo  {journal} {Quarterly Journal
				of Experimental Physiology: Translation and Integration}\ }\textbf {\bibinfo
			{volume} {22}},\ \bibinfo {pages} {281} (\bibinfo {year} {1932})}\BibitemShut
	{NoStop}%
	\bibitem [{\citenamefont {Kernick}\ \emph {et~al.}(1973)\citenamefont
		{Kernick}, \citenamefont {Jay}, \citenamefont {Rowlands},\ and\ \citenamefont
		{Skibo}}]{kernick1973experiments}%
	\BibitemOpen
	\bibfield  {author} {\bibinfo {author} {\bibfnamefont {D.}~\bibnamefont
			{Kernick}}, \bibinfo {author} {\bibfnamefont {A.}~\bibnamefont {Jay}},
		\bibinfo {author} {\bibfnamefont {S.}~\bibnamefont {Rowlands}},\ and\
		\bibinfo {author} {\bibfnamefont {L.}~\bibnamefont {Skibo}},\ }\bibfield
	{title} {\bibinfo {title} {Experiments on rouleau formation},\ }\href@noop {}
	{\bibfield  {journal} {\bibinfo  {journal} {Canadian Journal of Physiology
				and Pharmacology}\ }\textbf {\bibinfo {volume} {51}},\ \bibinfo {pages} {690}
		(\bibinfo {year} {1973})}\BibitemShut {NoStop}%
	\bibitem [{\citenamefont {Samsel}\ and\ \citenamefont
		{Perelson}(1982)}]{samsel1982kinetics}%
	\BibitemOpen
	\bibfield  {author} {\bibinfo {author} {\bibfnamefont {R.~W.}\ \bibnamefont
			{Samsel}}\ and\ \bibinfo {author} {\bibfnamefont {A.~S.}\ \bibnamefont
			{Perelson}},\ }\bibfield  {title} {\bibinfo {title} {Kinetics of rouleau
			formation. i. a mass action approach with geometric features},\ }\href@noop
	{} {\bibfield  {journal} {\bibinfo  {journal} {Biophysical Journal}\ }\textbf
		{\bibinfo {volume} {37}},\ \bibinfo {pages} {493} (\bibinfo {year}
		{1982})}\BibitemShut {NoStop}%
	\bibitem [{\citenamefont {Samsel}\ and\ \citenamefont
		{Perelson}(1984)}]{samsel1984kinetics}%
	\BibitemOpen
	\bibfield  {author} {\bibinfo {author} {\bibfnamefont {R.}~\bibnamefont
			{Samsel}}\ and\ \bibinfo {author} {\bibfnamefont {A.}~\bibnamefont
			{Perelson}},\ }\bibfield  {title} {\bibinfo {title} {Kinetics of rouleau
			formation. ii. reversible reactions},\ }\href@noop {} {\bibfield  {journal}
		{\bibinfo  {journal} {Biophysical journal}\ }\textbf {\bibinfo {volume}
			{45}},\ \bibinfo {pages} {805} (\bibinfo {year} {1984})}\BibitemShut
	{NoStop}%
	\bibitem [{\citenamefont {Perelson}\ and\ \citenamefont
		{Wiegel}(1982)}]{perelson1982equilibrium}%
	\BibitemOpen
	\bibfield  {author} {\bibinfo {author} {\bibfnamefont {A.~S.}\ \bibnamefont
			{Perelson}}\ and\ \bibinfo {author} {\bibfnamefont {F.~W.}\ \bibnamefont
			{Wiegel}},\ }\bibfield  {title} {\bibinfo {title} {The equilibrium size
			distribution of rouleaux},\ }\href@noop {} {\bibfield  {journal} {\bibinfo
			{journal} {Biophysical Journal}\ }\textbf {\bibinfo {volume} {37}},\ \bibinfo
		{pages} {515} (\bibinfo {year} {1982})}\BibitemShut {NoStop}%
	\bibitem [{\citenamefont {Bertoluzzo}\ \emph {et~al.}(1999)\citenamefont
		{Bertoluzzo}, \citenamefont {Bollini}, \citenamefont {Rasia},\ and\
		\citenamefont {Raynal}}]{bertoluzzo1999kinetic}%
	\BibitemOpen
	\bibfield  {author} {\bibinfo {author} {\bibfnamefont {S.}~\bibnamefont
			{Bertoluzzo}}, \bibinfo {author} {\bibfnamefont {A.}~\bibnamefont {Bollini}},
		\bibinfo {author} {\bibfnamefont {M.}~\bibnamefont {Rasia}},\ and\ \bibinfo
		{author} {\bibfnamefont {A.}~\bibnamefont {Raynal}},\ }\bibfield  {title}
	{\bibinfo {title} {Kinetic model for erythrocyte aggregation},\ }\href@noop
	{} {\bibfield  {journal} {\bibinfo  {journal} {Blood cells, molecules, and
				diseases}\ }\textbf {\bibinfo {volume} {25}},\ \bibinfo {pages} {339}
		(\bibinfo {year} {1999})}\BibitemShut {NoStop}%
	\bibitem [{\citenamefont {Babaki}\ \emph {et~al.}(2023)\citenamefont {Babaki},
		\citenamefont {Fedosov}, \citenamefont {Gholivand}, \citenamefont {Opdam},
		\citenamefont {Tuinier},\ and\ \citenamefont
		{Lettinga}}]{babaki2023competition}%
	\BibitemOpen
	\bibfield  {author} {\bibinfo {author} {\bibfnamefont {M.}~\bibnamefont
			{Babaki}}, \bibinfo {author} {\bibfnamefont {D.~A.}\ \bibnamefont {Fedosov}},
		\bibinfo {author} {\bibfnamefont {A.}~\bibnamefont {Gholivand}}, \bibinfo
		{author} {\bibfnamefont {J.}~\bibnamefont {Opdam}}, \bibinfo {author}
		{\bibfnamefont {R.}~\bibnamefont {Tuinier}},\ and\ \bibinfo {author}
		{\bibfnamefont {M.~P.}\ \bibnamefont {Lettinga}},\ }\bibfield  {title}
	{\bibinfo {title} {Competition between deformation and free volume quantified
			by 3d image analysis of red blood cell},\ }\href@noop {} {\bibfield
		{journal} {\bibinfo  {journal} {Biophysical Journal}\ }\textbf {\bibinfo
			{volume} {122}},\ \bibinfo {pages} {1646} (\bibinfo {year}
		{2023})}\BibitemShut {NoStop}%
	\bibitem [{\citenamefont {Ding}\ and\ \citenamefont
		{Aidun}(2006)}]{ding2006cluster}%
	\BibitemOpen
	\bibfield  {author} {\bibinfo {author} {\bibfnamefont {E.-J.}\ \bibnamefont
			{Ding}}\ and\ \bibinfo {author} {\bibfnamefont {C.~K.}\ \bibnamefont
			{Aidun}},\ }\bibfield  {title} {\bibinfo {title} {Cluster size distribution
			and scaling for spherical particles and red blood cells in pressure-driven
			flows at small reynolds number},\ }\href@noop {} {\bibfield  {journal}
		{\bibinfo  {journal} {Physical review letters}\ }\textbf {\bibinfo {volume}
			{96}},\ \bibinfo {pages} {204502} (\bibinfo {year} {2006})}\BibitemShut
	{NoStop}%
	\bibitem [{\citenamefont {Darras}\ \emph {et~al.}(2022)\citenamefont {Darras},
		\citenamefont {Dasanna}, \citenamefont {John}, \citenamefont {Gompper},
		\citenamefont {Kaestner}, \citenamefont {Fedosov},\ and\ \citenamefont
		{Wagner}}]{darras2022erythrocyte}%
	\BibitemOpen
	\bibfield  {author} {\bibinfo {author} {\bibfnamefont {A.}~\bibnamefont
			{Darras}}, \bibinfo {author} {\bibfnamefont {A.~K.}\ \bibnamefont {Dasanna}},
		\bibinfo {author} {\bibfnamefont {T.}~\bibnamefont {John}}, \bibinfo {author}
		{\bibfnamefont {G.}~\bibnamefont {Gompper}}, \bibinfo {author} {\bibfnamefont
			{L.}~\bibnamefont {Kaestner}}, \bibinfo {author} {\bibfnamefont {D.~A.}\
			\bibnamefont {Fedosov}},\ and\ \bibinfo {author} {\bibfnamefont
			{C.}~\bibnamefont {Wagner}},\ }\bibfield  {title} {\bibinfo {title}
		{Erythrocyte sedimentation: Collapse of a high-volume-fraction soft-particle
			gel},\ }\href@noop {} {\bibfield  {journal} {\bibinfo  {journal} {Physical
				Review Letters}\ }\textbf {\bibinfo {volume} {128}},\ \bibinfo {pages}
		{088101} (\bibinfo {year} {2022})}\BibitemShut {NoStop}%
	\bibitem [{\citenamefont {Olla}(1999)}]{olla1999simplified}%
	\BibitemOpen
	\bibfield  {author} {\bibinfo {author} {\bibfnamefont {P.}~\bibnamefont
			{Olla}},\ }\bibfield  {title} {\bibinfo {title} {Simplified model for red
			cell dynamics in small blood vessels},\ }\href@noop {} {\bibfield  {journal}
		{\bibinfo  {journal} {Physical review letters}\ }\textbf {\bibinfo {volume}
			{82}},\ \bibinfo {pages} {453} (\bibinfo {year} {1999})}\BibitemShut
	{NoStop}%
	\bibitem [{\citenamefont {Kaoui}\ \emph {et~al.}(2009)\citenamefont {Kaoui},
		\citenamefont {Biros},\ and\ \citenamefont {Misbah}}]{kaoui2009red}%
	\BibitemOpen
	\bibfield  {author} {\bibinfo {author} {\bibfnamefont {B.}~\bibnamefont
			{Kaoui}}, \bibinfo {author} {\bibfnamefont {G.}~\bibnamefont {Biros}},\ and\
		\bibinfo {author} {\bibfnamefont {C.}~\bibnamefont {Misbah}},\ }\bibfield
	{title} {\bibinfo {title} {Why do red blood cells have asymmetric shapes even
			in a symmetric flow?},\ }\href@noop {} {\bibfield  {journal} {\bibinfo
			{journal} {Physical review letters}\ }\textbf {\bibinfo {volume} {103}},\
		\bibinfo {pages} {188101} (\bibinfo {year} {2009})}\BibitemShut {NoStop}%
	\bibitem [{\citenamefont {Mauer}\ \emph {et~al.}(2018)\citenamefont {Mauer},
		\citenamefont {Mendez}, \citenamefont {Lanotte}, \citenamefont {Nicoud},
		\citenamefont {Abkarian}, \citenamefont {Gompper},\ and\ \citenamefont
		{Fedosov}}]{mauer2018flow}%
	\BibitemOpen
	\bibfield  {author} {\bibinfo {author} {\bibfnamefont {J.}~\bibnamefont
			{Mauer}}, \bibinfo {author} {\bibfnamefont {S.}~\bibnamefont {Mendez}},
		\bibinfo {author} {\bibfnamefont {L.}~\bibnamefont {Lanotte}}, \bibinfo
		{author} {\bibfnamefont {F.}~\bibnamefont {Nicoud}}, \bibinfo {author}
		{\bibfnamefont {M.}~\bibnamefont {Abkarian}}, \bibinfo {author}
		{\bibfnamefont {G.}~\bibnamefont {Gompper}},\ and\ \bibinfo {author}
		{\bibfnamefont {D.~A.}\ \bibnamefont {Fedosov}},\ }\bibfield  {title}
	{\bibinfo {title} {Flow-induced transitions of red blood cell shapes under
			shear},\ }\href@noop {} {\bibfield  {journal} {\bibinfo  {journal} {Physical
				review letters}\ }\textbf {\bibinfo {volume} {121}},\ \bibinfo {pages}
		{118103} (\bibinfo {year} {2018})}\BibitemShut {NoStop}%
	\bibitem [{\citenamefont {Gov}\ \emph {et~al.}(2003)\citenamefont {Gov},
		\citenamefont {Zilman},\ and\ \citenamefont {Safran}}]{gov2003cytoskeleton}%
	\BibitemOpen
	\bibfield  {author} {\bibinfo {author} {\bibfnamefont {N.}~\bibnamefont
			{Gov}}, \bibinfo {author} {\bibfnamefont {A.}~\bibnamefont {Zilman}},\ and\
		\bibinfo {author} {\bibfnamefont {S.}~\bibnamefont {Safran}},\ }\bibfield
	{title} {\bibinfo {title} {Cytoskeleton confinement and tension of red blood
			cell membranes},\ }\href@noop {} {\bibfield  {journal} {\bibinfo  {journal}
			{Physical review letters}\ }\textbf {\bibinfo {volume} {90}},\ \bibinfo
		{pages} {228101} (\bibinfo {year} {2003})}\BibitemShut {NoStop}%
	\bibitem [{\citenamefont {Rochal}\ and\ \citenamefont
		{Lorman}(2006)}]{rochal2006cytoskeleton}%
	\BibitemOpen
	\bibfield  {author} {\bibinfo {author} {\bibfnamefont {S.}~\bibnamefont
			{Rochal}}\ and\ \bibinfo {author} {\bibfnamefont {V.}~\bibnamefont
			{Lorman}},\ }\bibfield  {title} {\bibinfo {title} {Cytoskeleton influence on
			normal and tangent fluctuation modes in the red blood cells},\ }\href@noop {}
	{\bibfield  {journal} {\bibinfo  {journal} {Physical review letters}\
		}\textbf {\bibinfo {volume} {96}},\ \bibinfo {pages} {248102} (\bibinfo
		{year} {2006})}\BibitemShut {NoStop}%
	\bibitem [{\citenamefont {Sens}\ and\ \citenamefont
		{Gov}(2007)}]{sens2007force}%
	\BibitemOpen
	\bibfield  {author} {\bibinfo {author} {\bibfnamefont {P.}~\bibnamefont
			{Sens}}\ and\ \bibinfo {author} {\bibfnamefont {N.}~\bibnamefont {Gov}},\
	}\bibfield  {title} {\bibinfo {title} {Force balance and membrane shedding at
			the red-blood-cell surface},\ }\href@noop {} {\bibfield  {journal} {\bibinfo
			{journal} {Physical review letters}\ }\textbf {\bibinfo {volume} {98}},\
		\bibinfo {pages} {018102} (\bibinfo {year} {2007})}\BibitemShut {NoStop}%
	\bibitem [{\citenamefont {Boal}\ \emph {et~al.}(1992)\citenamefont {Boal},
		\citenamefont {Seifert},\ and\ \citenamefont {Zilker}}]{boal1992dual}%
	\BibitemOpen
	\bibfield  {author} {\bibinfo {author} {\bibfnamefont {D.~H.}\ \bibnamefont
			{Boal}}, \bibinfo {author} {\bibfnamefont {U.}~\bibnamefont {Seifert}},\ and\
		\bibinfo {author} {\bibfnamefont {A.}~\bibnamefont {Zilker}},\ }\bibfield
	{title} {\bibinfo {title} {Dual network model for red blood cell membranes},\
	}\href@noop {} {\bibfield  {journal} {\bibinfo  {journal} {Physical review
				letters}\ }\textbf {\bibinfo {volume} {69}},\ \bibinfo {pages} {3405}
		(\bibinfo {year} {1992})}\BibitemShut {NoStop}%
	\bibitem [{\citenamefont {Ben-Isaac}\ \emph {et~al.}(2011)\citenamefont
		{Ben-Isaac}, \citenamefont {Park}, \citenamefont {Popescu}, \citenamefont
		{Brown}, \citenamefont {Gov},\ and\ \citenamefont
		{Shokef}}]{ben2011effective}%
	\BibitemOpen
	\bibfield  {author} {\bibinfo {author} {\bibfnamefont {E.}~\bibnamefont
			{Ben-Isaac}}, \bibinfo {author} {\bibfnamefont {Y.}~\bibnamefont {Park}},
		\bibinfo {author} {\bibfnamefont {G.}~\bibnamefont {Popescu}}, \bibinfo
		{author} {\bibfnamefont {F.~L.}\ \bibnamefont {Brown}}, \bibinfo {author}
		{\bibfnamefont {N.~S.}\ \bibnamefont {Gov}},\ and\ \bibinfo {author}
		{\bibfnamefont {Y.}~\bibnamefont {Shokef}},\ }\bibfield  {title} {\bibinfo
		{title} {Effective temperature of red-blood-cell membrane fluctuations},\
	}\href@noop {} {\bibfield  {journal} {\bibinfo  {journal} {Physical review
				letters}\ }\textbf {\bibinfo {volume} {106}},\ \bibinfo {pages} {238103}
		(\bibinfo {year} {2011})}\BibitemShut {NoStop}%
	\bibitem [{\citenamefont {Pivkin}\ and\ \citenamefont
		{Karniadakis}(2008)}]{pivkin2008accurate}%
	\BibitemOpen
	\bibfield  {author} {\bibinfo {author} {\bibfnamefont {I.~V.}\ \bibnamefont
			{Pivkin}}\ and\ \bibinfo {author} {\bibfnamefont {G.~E.}\ \bibnamefont
			{Karniadakis}},\ }\bibfield  {title} {\bibinfo {title} {Accurate
			coarse-grained modeling of red blood cells},\ }\href@noop {} {\bibfield
		{journal} {\bibinfo  {journal} {Physical review letters}\ }\textbf {\bibinfo
			{volume} {101}},\ \bibinfo {pages} {118105} (\bibinfo {year}
		{2008})}\BibitemShut {NoStop}%
	\bibitem [{\citenamefont {Grandchamp}\ \emph {et~al.}(2013)\citenamefont
		{Grandchamp}, \citenamefont {Coupier}, \citenamefont {Srivastav},
		\citenamefont {Minetti},\ and\ \citenamefont
		{Podgorski}}]{grandchamp2013lift}%
	\BibitemOpen
	\bibfield  {author} {\bibinfo {author} {\bibfnamefont {X.}~\bibnamefont
			{Grandchamp}}, \bibinfo {author} {\bibfnamefont {G.}~\bibnamefont {Coupier}},
		\bibinfo {author} {\bibfnamefont {A.}~\bibnamefont {Srivastav}}, \bibinfo
		{author} {\bibfnamefont {C.}~\bibnamefont {Minetti}},\ and\ \bibinfo {author}
		{\bibfnamefont {T.}~\bibnamefont {Podgorski}},\ }\bibfield  {title} {\bibinfo
		{title} {Lift and down-gradient shear-induced diffusion in red blood cell
			suspensions},\ }\href@noop {} {\bibfield  {journal} {\bibinfo  {journal}
			{Physical review letters}\ }\textbf {\bibinfo {volume} {110}},\ \bibinfo
		{pages} {108101} (\bibinfo {year} {2013})}\BibitemShut {NoStop}%
	\bibitem [{\citenamefont {Shen}\ \emph {et~al.}(2018)\citenamefont {Shen},
		\citenamefont {Fischer}, \citenamefont {Farutin}, \citenamefont {Vlahovska},
		\citenamefont {Harting},\ and\ \citenamefont {Misbah}}]{shen2018blood}%
	\BibitemOpen
	\bibfield  {author} {\bibinfo {author} {\bibfnamefont {Z.}~\bibnamefont
			{Shen}}, \bibinfo {author} {\bibfnamefont {T.~M.}\ \bibnamefont {Fischer}},
		\bibinfo {author} {\bibfnamefont {A.}~\bibnamefont {Farutin}}, \bibinfo
		{author} {\bibfnamefont {P.~M.}\ \bibnamefont {Vlahovska}}, \bibinfo {author}
		{\bibfnamefont {J.}~\bibnamefont {Harting}},\ and\ \bibinfo {author}
		{\bibfnamefont {C.}~\bibnamefont {Misbah}},\ }\bibfield  {title} {\bibinfo
		{title} {Blood crystal: emergent order of red blood cells under wall-confined
			shear flow},\ }\href@noop {} {\bibfield  {journal} {\bibinfo  {journal}
			{Physical review letters}\ }\textbf {\bibinfo {volume} {120}},\ \bibinfo
		{pages} {268102} (\bibinfo {year} {2018})}\BibitemShut {NoStop}%
	\bibitem [{\citenamefont {Abkarian}\ \emph {et~al.}(2007)\citenamefont
		{Abkarian}, \citenamefont {Faivre},\ and\ \citenamefont
		{Viallat}}]{abkarian2007swinging}%
	\BibitemOpen
	\bibfield  {author} {\bibinfo {author} {\bibfnamefont {M.}~\bibnamefont
			{Abkarian}}, \bibinfo {author} {\bibfnamefont {M.}~\bibnamefont {Faivre}},\
		and\ \bibinfo {author} {\bibfnamefont {A.}~\bibnamefont {Viallat}},\
	}\bibfield  {title} {\bibinfo {title} {Swinging of red blood cells under
			shear flow},\ }\href@noop {} {\bibfield  {journal} {\bibinfo  {journal}
			{Physical review letters}\ }\textbf {\bibinfo {volume} {98}},\ \bibinfo
		{pages} {188302} (\bibinfo {year} {2007})}\BibitemShut {NoStop}%
	\bibitem [{\citenamefont {Dupire}\ \emph {et~al.}(2010)\citenamefont {Dupire},
		\citenamefont {Abkarian},\ and\ \citenamefont {Viallat}}]{dupire2010chaotic}%
	\BibitemOpen
	\bibfield  {author} {\bibinfo {author} {\bibfnamefont {J.}~\bibnamefont
			{Dupire}}, \bibinfo {author} {\bibfnamefont {M.}~\bibnamefont {Abkarian}},\
		and\ \bibinfo {author} {\bibfnamefont {A.}~\bibnamefont {Viallat}},\
	}\bibfield  {title} {\bibinfo {title} {Chaotic dynamics of red blood cells in
			a sinusoidal flow},\ }\href@noop {} {\bibfield  {journal} {\bibinfo
			{journal} {Physical review letters}\ }\textbf {\bibinfo {volume} {104}},\
		\bibinfo {pages} {168101} (\bibinfo {year} {2010})}\BibitemShut {NoStop}%
	\bibitem [{\citenamefont {Skotheim}\ and\ \citenamefont
		{Secomb}(2007)}]{skotheim2007red}%
	\BibitemOpen
	\bibfield  {author} {\bibinfo {author} {\bibfnamefont {J.}~\bibnamefont
			{Skotheim}}\ and\ \bibinfo {author} {\bibfnamefont {T.~W.}\ \bibnamefont
			{Secomb}},\ }\bibfield  {title} {\bibinfo {title} {Red blood cells and other
			nonspherical capsules in shear flow: oscillatory dynamics and the
			tank-treading-to-tumbling transition},\ }\href@noop {} {\bibfield  {journal}
		{\bibinfo  {journal} {Physical review letters}\ }\textbf {\bibinfo {volume}
			{98}},\ \bibinfo {pages} {078301} (\bibinfo {year} {2007})}\BibitemShut
	{NoStop}%
	\bibitem [{\citenamefont {Abbasi}\ \emph {et~al.}(2021)\citenamefont {Abbasi},
		\citenamefont {Farutin}, \citenamefont {Ez-Zahraouy}, \citenamefont
		{Benyoussef},\ and\ \citenamefont {Misbah}}]{abbasi2021erythrocyte}%
	\BibitemOpen
	\bibfield  {author} {\bibinfo {author} {\bibfnamefont {M.}~\bibnamefont
			{Abbasi}}, \bibinfo {author} {\bibfnamefont {A.}~\bibnamefont {Farutin}},
		\bibinfo {author} {\bibfnamefont {H.}~\bibnamefont {Ez-Zahraouy}}, \bibinfo
		{author} {\bibfnamefont {A.}~\bibnamefont {Benyoussef}},\ and\ \bibinfo
		{author} {\bibfnamefont {C.}~\bibnamefont {Misbah}},\ }\bibfield  {title}
	{\bibinfo {title} {Erythrocyte-erythrocyte aggregation dynamics under shear
			flow},\ }\href@noop {} {\bibfield  {journal} {\bibinfo  {journal} {Physical
				Review Fluids}\ }\textbf {\bibinfo {volume} {6}},\ \bibinfo {pages} {023602}
		(\bibinfo {year} {2021})}\BibitemShut {NoStop}%
	\bibitem [{\citenamefont {Ming}\ \emph {et~al.}(1965)\citenamefont {Ming},
		\citenamefont {Goodman},\ and\ \citenamefont {BROWN}}]{ming1965mathematical}%
	\BibitemOpen
	\bibfield  {author} {\bibinfo {author} {\bibfnamefont {T.~K.}\ \bibnamefont
			{Ming}}, \bibinfo {author} {\bibfnamefont {H.~S.}\ \bibnamefont {Goodman}},\
		and\ \bibinfo {author} {\bibfnamefont {B.}~\bibnamefont {BROWN}},\ }\bibfield
	{title} {\bibinfo {title} {Mathematical model for the process of aggregation
			in immune agglutination},\ }\href@noop {} {\bibfield  {journal} {\bibinfo
			{journal} {Nature}\ }\textbf {\bibinfo {volume} {208}},\ \bibinfo {pages}
		{84} (\bibinfo {year} {1965})}\BibitemShut {NoStop}%
	\bibitem [{\citenamefont {Fenech}\ \emph {et~al.}(2009)\citenamefont {Fenech},
		\citenamefont {Garcia}, \citenamefont {Meiselman},\ and\ \citenamefont
		{Cloutier}}]{fenech2009particle}%
	\BibitemOpen
	\bibfield  {author} {\bibinfo {author} {\bibfnamefont {M.}~\bibnamefont
			{Fenech}}, \bibinfo {author} {\bibfnamefont {D.}~\bibnamefont {Garcia}},
		\bibinfo {author} {\bibfnamefont {H.~J.}\ \bibnamefont {Meiselman}},\ and\
		\bibinfo {author} {\bibfnamefont {G.}~\bibnamefont {Cloutier}},\ }\bibfield
	{title} {\bibinfo {title} {A particle dynamic model of red blood cell
			aggregation kinetics},\ }\href@noop {} {\bibfield  {journal} {\bibinfo
			{journal} {Annals of biomedical engineering}\ }\textbf {\bibinfo {volume}
			{37}},\ \bibinfo {pages} {2299} (\bibinfo {year} {2009})}\BibitemShut
	{NoStop}%
	\bibitem [{\citenamefont {Nehring}\ \emph {et~al.}(2018)\citenamefont
		{Nehring}, \citenamefont {Shendruk},\ and\ \citenamefont
		{de~Haan}}]{nehring2018morphology}%
	\BibitemOpen
	\bibfield  {author} {\bibinfo {author} {\bibfnamefont {A.}~\bibnamefont
			{Nehring}}, \bibinfo {author} {\bibfnamefont {T.~N.}\ \bibnamefont
			{Shendruk}},\ and\ \bibinfo {author} {\bibfnamefont {H.~W.}\ \bibnamefont
			{de~Haan}},\ }\bibfield  {title} {\bibinfo {title} {Morphology of
			depletant-induced erythrocyte aggregates},\ }\href@noop {} {\bibfield
		{journal} {\bibinfo  {journal} {Soft matter}\ }\textbf {\bibinfo {volume}
			{14}},\ \bibinfo {pages} {8160} (\bibinfo {year} {2018})}\BibitemShut
	{NoStop}%
	\bibitem [{\citenamefont {Bak}\ \emph {et~al.}(1988)\citenamefont {Bak},
		\citenamefont {Tang},\ and\ \citenamefont {Wiesenfeld}}]{bak1988self}%
	\BibitemOpen
	\bibfield  {author} {\bibinfo {author} {\bibfnamefont {P.}~\bibnamefont
			{Bak}}, \bibinfo {author} {\bibfnamefont {C.}~\bibnamefont {Tang}},\ and\
		\bibinfo {author} {\bibfnamefont {K.}~\bibnamefont {Wiesenfeld}},\ }\bibfield
	{title} {\bibinfo {title} {Self-organized criticality},\ }\href@noop {}
	{\bibfield  {journal} {\bibinfo  {journal} {Physical review A}\ }\textbf
		{\bibinfo {volume} {38}},\ \bibinfo {pages} {364} (\bibinfo {year}
		{1988})}\BibitemShut {NoStop}%
	\bibitem [{\citenamefont {Bak}\ \emph {et~al.}()\citenamefont {Bak},
		\citenamefont {Tang},\ and\ \citenamefont {Wiesenfeld}}]{bak59self}%
	\BibitemOpen
	\bibfield  {author} {\bibinfo {author} {\bibfnamefont {P.}~\bibnamefont
			{Bak}}, \bibinfo {author} {\bibfnamefont {C.}~\bibnamefont {Tang}},\ and\
		\bibinfo {author} {\bibfnamefont {K.}~\bibnamefont {Wiesenfeld}},\ }\bibfield
	{title} {\bibinfo {title} {Self-organized criticality: an explanation of 1/f
			noise, 1987},\ }\href@noop {} {\bibfield  {journal} {\bibinfo  {journal}
			{Phys. Rev. Lett}\ }\textbf {\bibinfo {volume} {59}},\ \bibinfo {pages}
		{381}}\BibitemShut {NoStop}%
	\bibitem [{\citenamefont {Lee}\ \emph {et~al.}(2016)\citenamefont {Lee},
		\citenamefont {Kinnunen}, \citenamefont {Khokhlova}, \citenamefont {Lyubin},
		\citenamefont {Priezzhev}, \citenamefont {Meglinski},\ and\ \citenamefont
		{Fedyanin}}]{lee2016optical}%
	\BibitemOpen
	\bibfield  {author} {\bibinfo {author} {\bibfnamefont {K.}~\bibnamefont
			{Lee}}, \bibinfo {author} {\bibfnamefont {M.}~\bibnamefont {Kinnunen}},
		\bibinfo {author} {\bibfnamefont {M.~D.}\ \bibnamefont {Khokhlova}}, \bibinfo
		{author} {\bibfnamefont {E.~V.}\ \bibnamefont {Lyubin}}, \bibinfo {author}
		{\bibfnamefont {A.~V.}\ \bibnamefont {Priezzhev}}, \bibinfo {author}
		{\bibfnamefont {I.}~\bibnamefont {Meglinski}},\ and\ \bibinfo {author}
		{\bibfnamefont {A.~A.}\ \bibnamefont {Fedyanin}},\ }\bibfield  {title}
	{\bibinfo {title} {Optical tweezers study of red blood cell aggregation and
			disaggregation in plasma and protein solutions},\ }\href@noop {} {\bibfield
		{journal} {\bibinfo  {journal} {Journal of biomedical optics}\ }\textbf
		{\bibinfo {volume} {21}},\ \bibinfo {pages} {035001} (\bibinfo {year}
		{2016})}\BibitemShut {NoStop}%
	\bibitem [{\citenamefont {Gisiger}(2001)}]{gisiger2001scale}%
	\BibitemOpen
	\bibfield  {author} {\bibinfo {author} {\bibfnamefont {T.}~\bibnamefont
			{Gisiger}},\ }\bibfield  {title} {\bibinfo {title} {Scale invariance in
			biology: coincidence or footprint of a universal mechanism?},\ }\href@noop {}
	{\bibfield  {journal} {\bibinfo  {journal} {Biological Reviews}\ }\textbf
		{\bibinfo {volume} {76}},\ \bibinfo {pages} {161} (\bibinfo {year}
		{2001})}\BibitemShut {NoStop}%
	\bibitem [{\citenamefont {Kinnunen}\ \emph {et~al.}(2011)\citenamefont
		{Kinnunen}, \citenamefont {Kauppila}, \citenamefont {Karmenyan},\ and\
		\citenamefont {Myllyl{\"a}}}]{kinnunen2011effect}%
	\BibitemOpen
	\bibfield  {author} {\bibinfo {author} {\bibfnamefont {M.}~\bibnamefont
			{Kinnunen}}, \bibinfo {author} {\bibfnamefont {A.}~\bibnamefont {Kauppila}},
		\bibinfo {author} {\bibfnamefont {A.}~\bibnamefont {Karmenyan}},\ and\
		\bibinfo {author} {\bibfnamefont {R.}~\bibnamefont {Myllyl{\"a}}},\
	}\bibfield  {title} {\bibinfo {title} {Effect of the size and shape of a red
			blood cell on elastic light scattering properties at the single-cell level},\
	}\href@noop {} {\bibfield  {journal} {\bibinfo  {journal} {Biomedical optics
				express}\ }\textbf {\bibinfo {volume} {2}},\ \bibinfo {pages} {1803}
		(\bibinfo {year} {2011})}\BibitemShut {NoStop}%
	\bibitem [{\citenamefont {Phillips}\ \emph {et~al.}(2012)\citenamefont
		{Phillips}, \citenamefont {Jacques},\ and\ \citenamefont
		{McCarty}}]{phillips2012measurement}%
	\BibitemOpen
	\bibfield  {author} {\bibinfo {author} {\bibfnamefont {K.~G.}\ \bibnamefont
			{Phillips}}, \bibinfo {author} {\bibfnamefont {S.~L.}\ \bibnamefont
			{Jacques}},\ and\ \bibinfo {author} {\bibfnamefont {O.~J.}\ \bibnamefont
			{McCarty}},\ }\bibfield  {title} {\bibinfo {title} {Measurement of single
			cell refractive index, dry mass, volume, and density using a
			transillumination microscope},\ }\href@noop {} {\bibfield  {journal}
		{\bibinfo  {journal} {Physical review letters}\ }\textbf {\bibinfo {volume}
			{109}},\ \bibinfo {pages} {118105} (\bibinfo {year} {2012})}\BibitemShut
	{NoStop}%
\end{thebibliography}
%\bibliographystyle{aps}

\end{document}